\documentclass{article}

\usepackage[dvips]{graphicx}
\usepackage{cite}

\newtheorem{dfn}{Definition}

\title{A fully quantum method of determination of penetrability and reflection coefficients in quantum FRW model with radiation}

\author{Sergei~P.~Maydanyuk%
\thanks{\emph{E-mail:} maidan@kinr.kiev.ua} \\
\small\emph{Institute for Nuclear Research, National Academy of Sciences of Ukraine} \\
\small\emph{47, prosp. Nauki, Kiev-28, 03680, Ukraine}}
\date{\small\today}

\begin{document}

\maketitle

\begin{abstract}
In the paper the closed Friedmann--Robertson--Walker model with quantization in the presence of the positive cosmological constant and radiation is studied. For analysis of tunneling probability for birth of an asymptotically deSitter, inflationary Universe as a function of the radiation energy a new definition of a ``free'' wave propagating inside strong fields is proposed.
On such a basis, tunneling boundary condition is corrected, penetrability and reflection concerning to the barrier are calculated in fully quantum stationary approach. For the first time non-zero interference between the incident and reflected waves has been taken into account which turns out to play important role inside cosmological potentials and could be explained by non-locality of barriers in quantum mechanics.
Inside whole region of energy of radiation the tunneling probability for the birth of the inflationary Universe is found to be close to its value obtained in semiclassical approach.
The reflection from the barrier is determined for the first time (which is essentially differs on 1 at the energy of radiation close to the barrier height).
The proposed method could be easily generalized on the cosmological models with the barriers of arbitrary shape, that has been demonstrated for the FRW-model with included Chaplygin gas.
Result is stable for variations of the studied barriers, accuracy are found to be 11--18 digits for all coefficients and energies below the barrier height.
\end{abstract}


\textbf{PACS nubers:}
98.80.Qc, 
98.80.–k, 
98.80.Bp, 
98.80.Jk, 
03.65.Xp 


\section{Introduction
\label{sec.introduction}}

In order to understand what really happens in the formation of the Universe, many people came to the point of view that a quantum consideration of this process is deeper. The first papers with the quantum approach for the description of Universe formation and its initial expansion may be~\cite{DeWitt.1967,Wheeler.1968}, and shortly afterwards many other papers appeared in this field, pointing to a rapid development of the quantum approach in cosmology (for example, see the first some papers \cite{Vilenkin.1982.PLB,Hartle.1983.PRD,Linde.1984.LNC,Zeldovich.1984.LNC,Rubakov.1984.PLB,Vilenkin.1984.PRD,Vilenkin.1986.PRD,Atkatz.1984.PRD} and some discussions in Refs.~\cite{Vilenkin.1994.PRD,Rubakov.1999} with references therein).

Today, among all variety of models one can select two approaches which are the most prevailing: these are the Feynman formalism of path integrals in multidimensional spacetime, developed by the Cambridge group and other researchers, called the \emph{``Hartle--Hawking method''} (for example, see Ref.~\cite{Hartle.1983.PRD}), and a method based on direct consideration of tunneling in 4-dimensional Euclidian spacetime, called the \emph{``Vilenkin method''}
(for example, see Refs.~\cite{Vilenkin.1982.PLB,Vilenkin.1984.PRD,Vilenkin.1986.PRD,Vilenkin.1994.PRD}). Here, according to Ref.~\cite{Vilenkin.1995}, in the quantum approach we have the following picture of the Universe creation: a closed Universe with a small size is formed from ``nothing'' (vacuum), where by the word ``nothing'' one refers to a quantum state without classical space and time. A wave function is used for a probabilistic description of the creation of the Universe and such a process is connected with transition of a wave through an effective barrier. Determination of penetrability of this barrier is a key point in estimation of duration of the formation of the Universe, and dynamics of its expansion in the first stage.

However, in majority of these models (with the exception of some exactly solvable models) tunneling is practically studied in details in the semiclassical approximation mainly (for example, see Refs.~\cite{Vilenkin.1994.PRD,Rubakov.1999}). An attractive side of such an approach is its simplicity in the construction of decreasing and increasing partial solutions for the wave function in the tunneling region, an outgoing wave function in the external region, and a possibility to define and to estimate in a simply enough way a penetrability of the barrier, which can be used for obtaining the duration of the nucleation of the Universe. A \emph{tunneling boundary condition} \cite{Vilenkin.1995,Vilenkin.1994.PRD} could seems to be the most natural and clear, where the wave function should represent an outgoing wave only in the enough large value of the scale factor $a$. However, whether is such a wave really free in the asymptotic region? If to draw attention on increasing of a modulus of the potential with increasing of the scale factor $a$ and increasing of a gradient of such a potential, used with opposite sign and having a sense of force, acting ``through the barrier'' on this wave, then one come to a serious contradiction: \emph{influence of the potential on this wave increases strongly at increasing of $a$}!
Now a new question has been appeared: what should the wave represent in a general in the cosmological problem?
This problem connects with another (new) else more general one in general quantum physics --- a real importance \emph{to define a ``free'' wave inside \underline{strong} fields} and we need in mathematical stable tools for its study and work. It becomes unclear whether a connection between exact solutions for the wave function at turning point and ``free'' wave defined in the asymptotic region is corrected.
If in frameworks of semiclassical approach (up to the second order) it is corrected then any deformations of tail of the potential after the barrier cannot change the penetrability absolutely, while force mentioned above and acting on the asymptotic ``free'' wave could be increased up to infinity! Answer on such misunderstanding could be found in non-locality of definition of the penetrability in quantum mechanics, which is reduced to minimum in the semiclassical approach (i.~e. this is so called ``error'' of the cosmological semiclassical approach).
Now, if to refuse application of the semiclassical approximation, then we come to a necessity to define such a wave as maximal correctly and accurately as a possible. The problem of the correct definition of the wave in cosmology is reinforced else more, if to calculate incident and reflected waves in the internal region. \emph{Even with the known exact solution for the wave function there is uncertainty in determination of these waves!} But, namely, the standard definition of the coefficients of penetrability and reflection is based on them. In particular, I have not found papers yet where the coefficient of reflection is defined and estimated in this problem (which differs essentially from unity at the energy of radiation close to height of the barrier and, therefore, such a characteristics could be interesting from a physical point of view). Note that the semiclassical approximation prejudices a possibility of its definition at all \cite{Landau.v3.1989}.

Thus, in order to estimate probability of the formation of the Universe as accurately as possible, we need in the fully quantum definition of the wave. Note that the non-semiclassical penetrability of the barrier in the cosmological problems has not been studied in details and, therefore, a development of fully quantum methods for its estimation is a perspective task. I would like to note undoubtedly perspective direction of researches~\cite{AcacioDeBarros.2007.PRD}, where the penetrability was estimated on the basis of tunneling of wave packet through the barrier. However, a stationary boundary condition has uncertainty that could lead to different results in calculations of the penetrability. The stationary approach could allow to clarify this deeply. It is able to give stale solutions for the wave function (and results in Ref.~\cite{Maydanyuk.2008.EPJC} have confirmed this at zero energy of radiation), uses the standard definition of the coefficients of the penetrability and reflection, is more accurate in their estimation.

Aims of this paper are:
(1) to define the wave in the quantum cosmological problem;
(2) to construct the fully quantum (non-semiclassical) stationary method of determination of the coefficients of penetrability of the barriers and reflection from them on the basis of such a definition of the wave;
(3) to estimate how much the semiclassical approach is differed in the estimation of the penetrability from the fully quantum one.
In order to achieve this, at first it needs to construct tools for stable calculation of partial solutions of the wave function. In order to resolve the questions pointed out above, we shall restrict ourselves by a simple cosmological model, where the potential has a barrier and internal above-barrier region.
The paper is organized so. After a short description of closed Friedmann--Robertson--Walker (FRW) model with quantization in a presence of a positive cosmological constant and radiation in Sec.~2, in Sec.~3 a new formalism for calculation of two linear independent partial solutions for the wave function of the Universe for the scale factor inside the region $0 \le a \le 100$ and the energy of radiation from zero up to the barrier height is presented.
In Sec.~4 a fully quantum definition of the wave is formulated (for the first time), a new approach for determination of the incident, reflected and transmitted waves relatively the barrier is constructed on such basis, the boundary condition is corrected.
In Sec.~5 a fully quantum stationary method of determination of coefficients of penetrability and reflection relatively the barrier with analysis of uniqueness of solution is presented, where at first time non-zero interference between the incident and reflected waves is analyzed and for its estimation the coefficient of mixing is introduced.
In such an approach the penetrability and reflection for the barrier for the studied cosmological model with $A=36$, $B=12\,\Lambda$ parameters at $\Lambda=0.01$ are calculated with estimation of accuracy and in comparison with the semiclassical results and with results of non-stationary quantum approach \cite{AcacioDeBarros.2007.PRD}.
In Sec.~6 the penetrability is estimated for the FRW-model with included Chaplygin gas which has internal hole before the barrier.
Conclusions finalize the paper.


\section{Cosmological model in the Friedmann--Robertson--Walker metric with radiation
\label{sec.model}}


Let us start from a case of a closed ($k=1$) FRW model in the presence of a positive cosmological constant $\Lambda > 0$ and radiation. The minisuperspace Lagrangian has the form
(see Appendix A, also Ref.~\cite{Vilenkin.1995}, (11), p.~4):
\begin{equation}
\begin{array}{lcl}
  \mathcal{L}\,(a,\dot{a}) =
  \displaystyle\frac{3\,a}{8\pi\,G}\:
  \biggl(-\dot{a}^{2} + k - \displaystyle\frac{8\pi\,G}{3}\; a^{2}\,\rho(a) \biggr), &
  \rho\,(a) = \rho_{\Lambda} + \displaystyle\frac{\rho_{\rm rad}}{a^{4}(t)}, &
  \rho_{\Lambda} = \displaystyle\frac{\Lambda}{8\pi\,G},
\end{array}
\label{eq.model.3.8}
\end{equation}
where
$a$ is scale factor,
$\dot{a}$ is derivative $a$ with respect to time coordinate $t$,
$\rho\,(a)$ is a general expression for the energy density,
$\rho_{\rm rad}(a)$ is component describing the radiation in the initial stage (equation of state for radiation is $p(a)=\rho_{\rm rad}(a)/3$, $p$ is pressure).
The passage to the quantum description of the evolution of the Universe is obtained by the standard procedure of canonical quantization in the Dirac formalism for systems with constraints. In result, we obtain the \emph{Wheeler--De Witt (WDW) equation} (see Ref.~\cite{Vilenkin.1995}, also \cite{Wheeler.1968,DeWitt.1967,Rubakov.2002.PRD}), which after multiplication on factor and passage of the item at the component with radiation $\rho_{\rm rad}$
into right part transforms into the following form (see Appendix A):
\begin{equation}
\begin{array}{ccl}
  \biggl\{ -\:\displaystyle\frac{\partial^{2}}{\partial a^{2}} + V\,(a) \biggr\}\; \varphi(a) =
  E_{\rm rad}\; \varphi(a), &
  V\, (a) =
    \biggl( \displaystyle\frac{3}{4\pi\,G} \biggr)^{2}\: k\,a^{2} -
    \displaystyle\frac{3\,\rho_{\Lambda}}{2\pi\,G}\; a^{4}, &
  E_{\rm rad} = \displaystyle\frac{3\,\rho_{\rm rad}}{2\pi\,G},
\end{array}
\label{eq.model.5.3}
\end{equation}
where $\varphi(a)$ is wave function of Universe. This equation looks similar to the one-dimensional stationary Schr\"{o}dinger equation on semiaxis (of the variable $a$) at energy $E_{\rm rad}$ with potential $V\,(a)$. For further analysis it is convenient to use the system of units where $8\pi\,G \equiv M_{\rm p} = 1$, and
to rewrite $V\,(a)$ in a generalized form:
\begin{equation}
  V(a) = A\,a^{2} - B\,a^{4}.
\label{eq.model.5.9}
\end{equation}
In particular, for the Universe of the closed type ($k=1$) we obtain $A = 36$, $B = 12\,\Lambda$ (this potential coincides with \cite{AcacioDeBarros.2007.PRD}).


In order to find wave function we shall needs to know shape of the potential close to turning points. Let us find the \emph{turning points} $a_{\rm tp,\,in}$ and $a_{\rm tp,\,out}$
concerning the potential (\ref{eq.model.5.9}) at energy $E_{\rm rad}$:
\begin{equation}
\begin{array}{cc}
\vspace{3mm}
  a_{\rm tp,\, in} =
    \sqrt{\displaystyle\frac{A}{2B}} \cdot
    \sqrt{1 - \sqrt{1 - \displaystyle\frac{4BE_{\rm rad}}{A^{2}}}},&
  a_{\rm tp,\, out} =
    \sqrt{\displaystyle\frac{A}{2B}} \cdot
    \sqrt{1 + \sqrt{1 - \displaystyle\frac{4BE_{\rm rad}}{A^{2}}}}.
\end{array}
\label{eq.model.6.2}
\end{equation}
Let us expand the potential $V(a)$ (\ref{eq.model.5.9}) in powers $q_{\rm out}=a-a_{\rm tp}$ (where as $a_{\rm tp}$ the point $a_{\rm tp,\, in}$ or $a_{\rm tp,\, out}$ is used, close to which we find expansion),
where (for small $q$) we restrict ourselves by the linear item only:
\begin{equation}
  V(q) = V_{0} + V_{1}\,q,
\label{eq.model.6.3}
\end{equation}
where the coefficients $V_{0}$ and $V_{1}$ are:
\begin{equation}
\begin{array}{lcl}
\vspace{1mm}
  V_{0} & = &
    V(a=a_{\rm tp,\, in}) =
    V(a=a_{\rm tp,\, out}) =
    A\, a_{\rm tp}^{2} - B\, a_{\rm tp}^{4} = E_{\rm rad}, \\
\vspace{1mm}
  V_{1}^{\rm (out)} & = &
    -\: 2\, A \cdot
    \sqrt{\displaystyle\frac{A}{2B}\:
    \biggl(1 - \displaystyle\frac{4BE_{\rm rad}}{A^{2}}\biggr)\,
    \biggl(1 + \sqrt{1 - \displaystyle\frac{4BE_{\rm rad}}{A^{2}}}\biggr)}, \\
  V_{1}^{\rm (int)} & = &
    2\,A \cdot
    \sqrt{\displaystyle\frac{A}{2B}\:
    \biggl(1 - \displaystyle\frac{4BE_{\rm rad}}{A^{2}}\biggr)\,
    \biggl(1 - \sqrt{1 - \displaystyle\frac{4BE_{\rm rad}}{A^{2}}}\biggr)}.
\end{array}
\label{eq.model.6.5}
\end{equation}
Now eq.~(\ref{eq.model.5.3}) transforms into a new form at variable $q$ with potential $V(q)$:
\begin{equation}
  -\displaystyle\frac{d^{2}}{dq^{2}}\, \varphi(q) +
  V_{1}\, q\: \varphi(q) = 0.
\label{eq.model.6.6}
\end{equation}

\section{Wave function: its behavior and partial solutions
\label{sec.3}}

The wave function is known to oscillate above the barrier and increase (or decrease) under the barrier without any oscillations. So, in order to provide a linear independence between two partial solutions for the wave function effectively, I look for the first partial solution increasing in the region of tunneling and the second one decreasing in this tunneling region. At first, I define each partial solution and its derivative at a selected starting point, and then I calculate them in the region close enough to this point using the \emph{method of beginning of the solution} presented in Appendix~\ref{sec.3.1}. Here, for the partial solution which increases in the barrier region, as the starting point I use internal turning point $a_{\rm tp,\, in}$ at non-zero energy $E_{\rm rad}$ or equals to zero $a=0$ at null energy $E_{\rm rad}$, and for the second partial solution which decreases in the barrier region, I select the starting point to be equal to external turning point $a_{\rm tp,\, out}$. Then both partial solutions and their derivatives I calculate independently in the whole required range of $a$ using the \emph{method of continuation of the solution} presented in Appendix~\ref{sec.3.2}, which is improvement of the Numerov method with a constant step. By such a way, I obtain two partial solutions for the wave function and their derivatives in the whole studied region.

\begin{figure}[h]
\centerline{
\includegraphics[width=55mm]{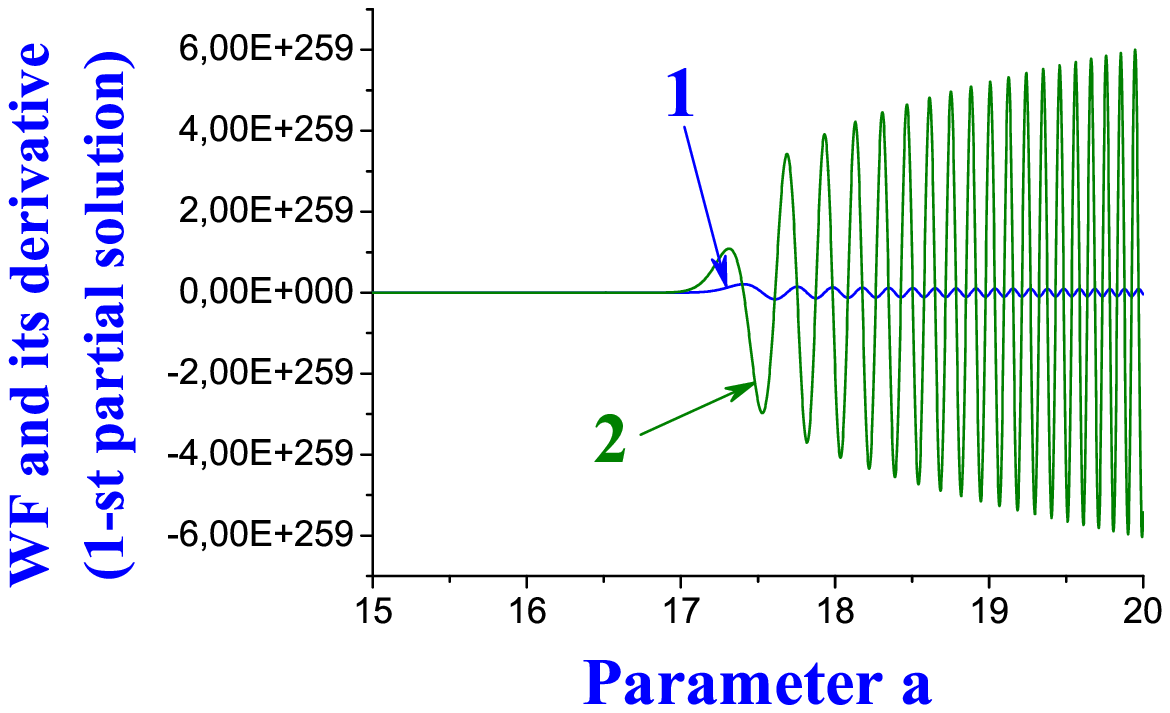}
\includegraphics[width=55mm]{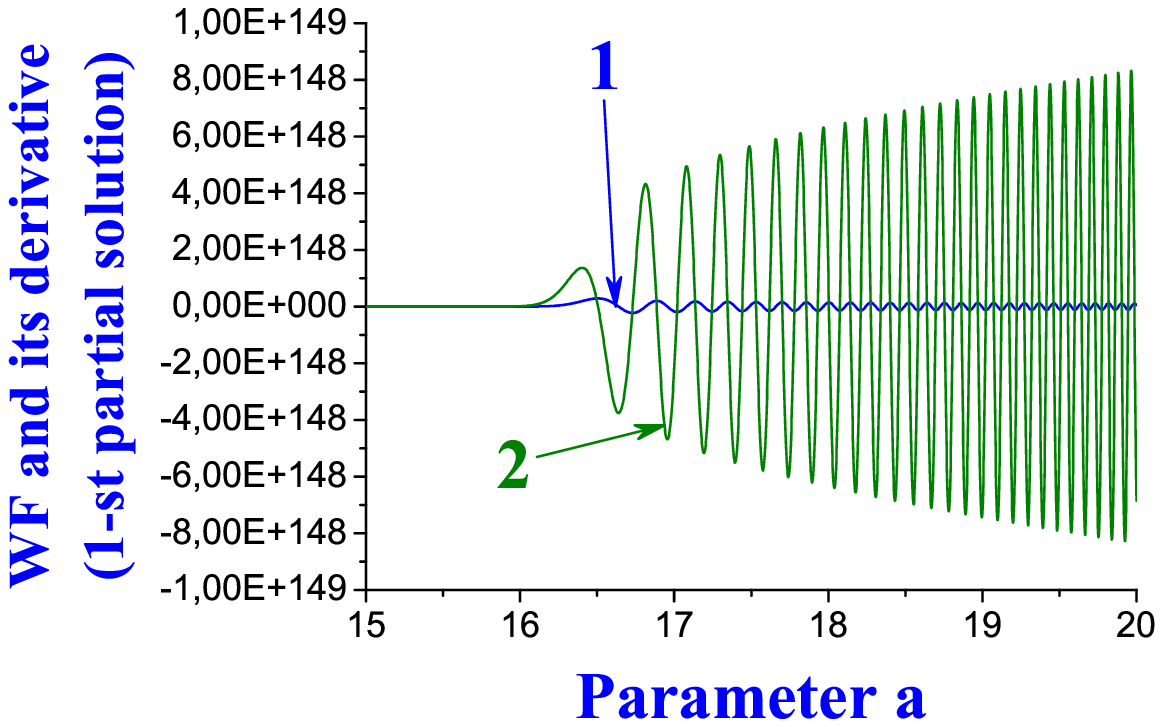}
\includegraphics[width=55mm]{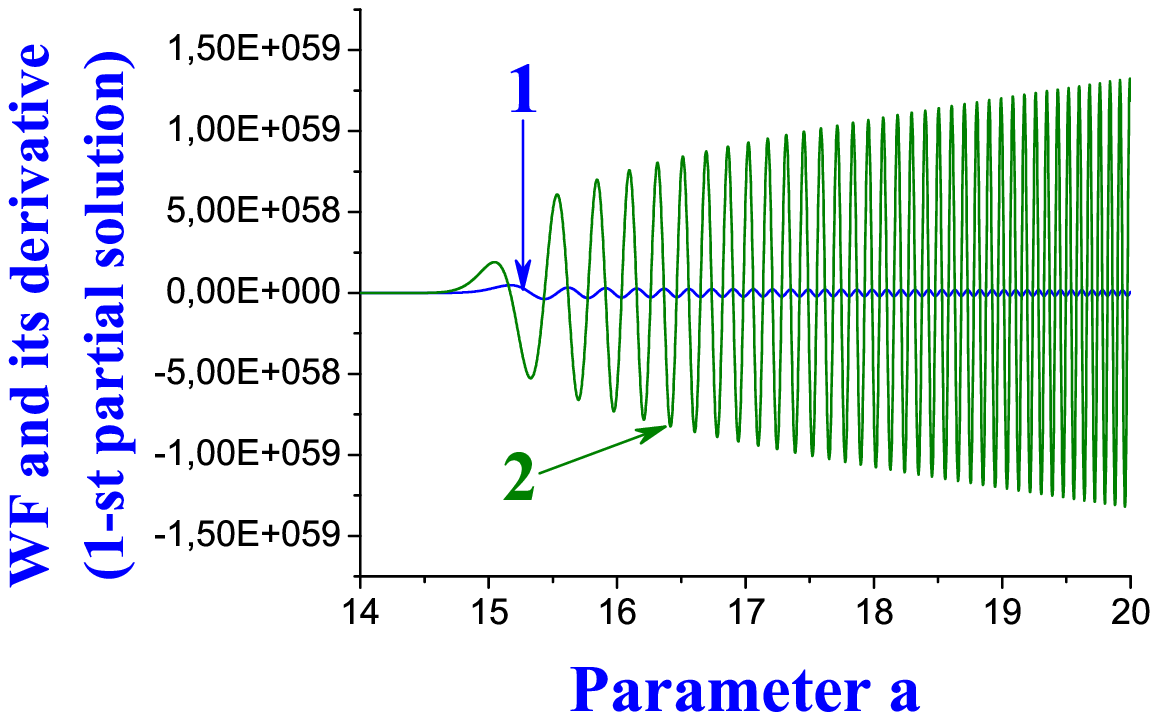}}
\caption{\small The first partial solution for the wave function and its derivative at different values of the energy of radiation $E_{\rm rad}$, increasing in the tunneling region (curve 1, blue, is for the wave function; curve 2, green, for the derivative of this wave function):
(a) $E_{\rm rad}=10$; (b) $E_{\rm rad}=1000$; (c) $E_{\rm rad}=2000$
\label{fig.2}}
\end{figure}

\begin{figure}[h]
\centerline{
\includegraphics[width=55mm]{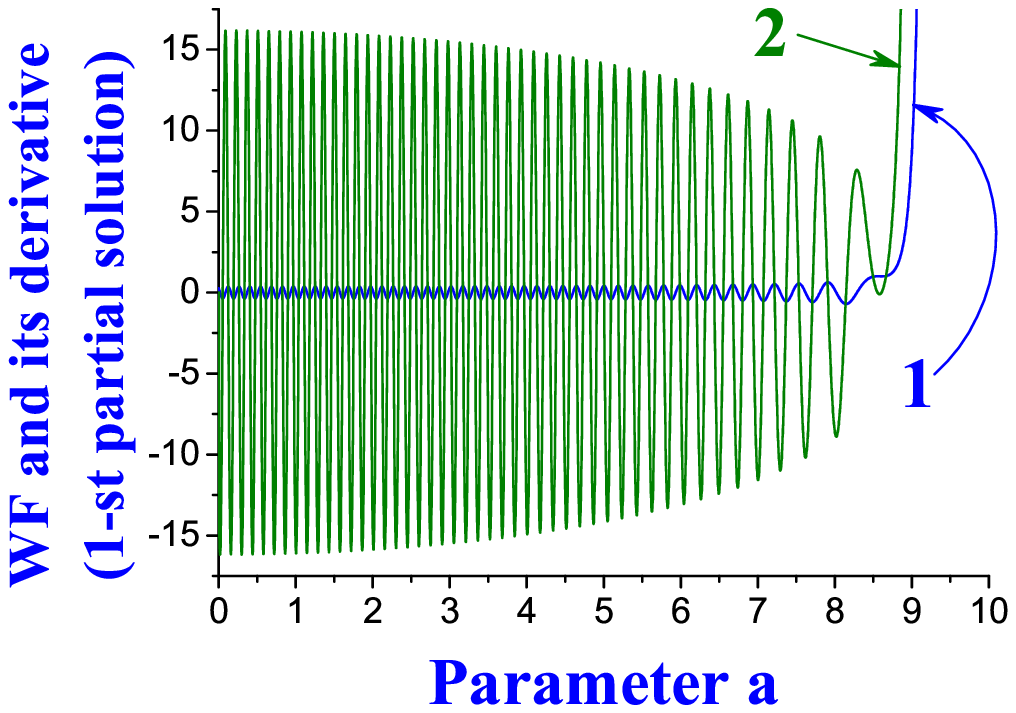}
\includegraphics[width=55mm]{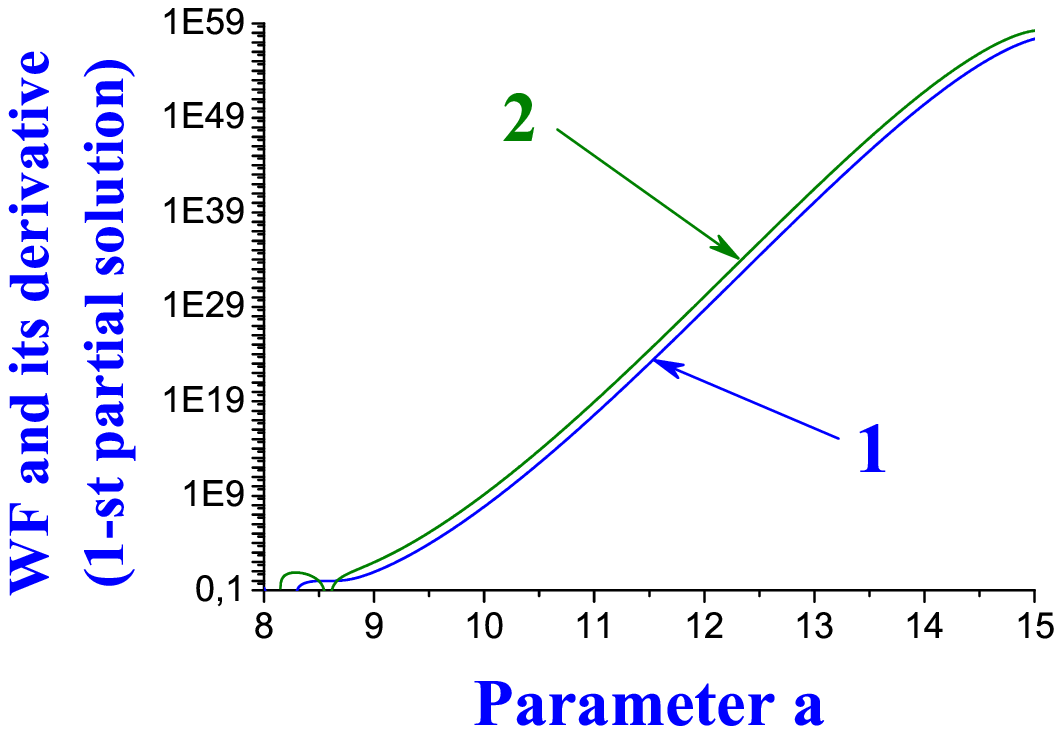}
\includegraphics[width=55mm]{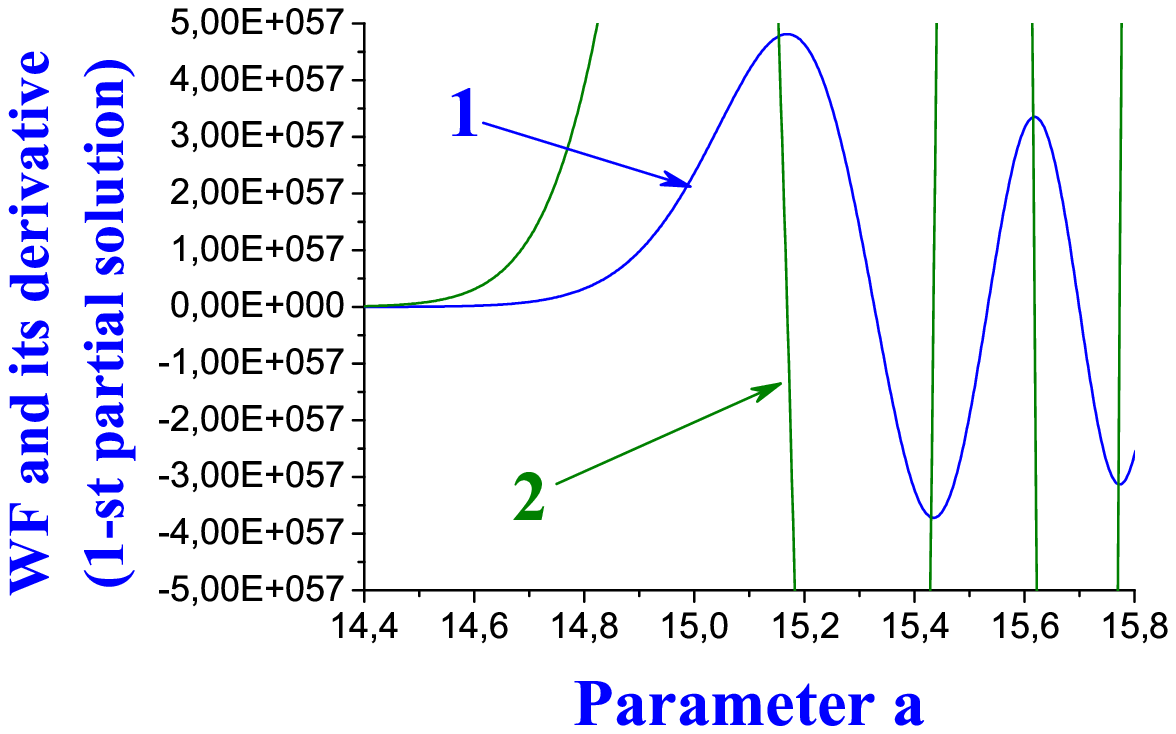}}
\caption{\small The first partial solution for the wave function and its derivative at the energy of radiation $E_{\rm rad}=2000$ (curve 1, blue, is for the wave function; curve 2, green, for the derivative of this wave function)
\label{fig.3}}
\end{figure}

In order to clarify how much the proposed approach gives convergent (stable) solutions, for a comparison we shall use paper~\cite{AcacioDeBarros.2007.PRD} with the published wave function (see (9) p.~5, I use the potential in eq.~(\ref{eq.model.5.9})).
Let us consider a behavior of the wave function. The first partial solution for the wave function and its derivative in my calculation are presented in Fig.~\ref{fig.2}, which increase in the tunneling region and
have been obtained at different values of the energy of radiation $E_{\rm rad}$.
Form these figures one can see that the wave function satisfies the rules of behavior of the wave functions inside sub-barrier and above-barrier regions \cite{Zakhariev.1990.PEPAN}. Starting from very small $a$, the wave function has oscillations and its maximums increase monotonously with increasing of $a$, which corresponds to behavior of the wave function in the internal region before the barrier (this becomes more obvious after essential increasing of scale, see left panel in Fig.~\ref{fig.3}). Further, with increasing of $a$ the wave function increases monotonously without any oscillations, that points out on transition into the tunneling region (one can see this in a logarithmic presentation of the wave function, see central panel in Fig.~\ref{fig.3}). A boundary of such a transformation in behavior of the wave function must be the point of penetration of the wave into the barrier, i.~e. the internal turning point  $a_{\rm tp,\,in}$. Further, with increasing of $a$ the oscillations are appeared in the wave function, which could be possible inside the above barrier region only (in the right panel of Fig.~\ref{fig.3} one can see that such a transition is extremely smooth that characterizes the accuracy of the method positively). A boundary of such a new transformation in the behavior of the wave function must be the point of leaving of the wave from the barrier outside $a_{\rm tp,\,out}$. Like Ref.~\cite{Maydanyuk.2008.EPJC}, but at arbitrary non-zero energy $E_{\rm rad}$ in the external region, starting from $a_{\rm tp,\,out}$, I obtain again a monotonous increasing of maximums of the derivative of the wave function and a smooth decreasing of this wave function. One can see that the derivative is larger essentially than the wave function. At enough large values of $a$, i.~e. in the region, which can be called \emph{asymptotic} one, I obtain the smooth continuous solutions,
achieving $a=100$ (in Ref.~\cite{AcacioDeBarros.2007.PRD} the maximal value is $a=30$).

The second partial solution of the wave function and its derivative in my calculation at different values of the energy of radiation $E_{\rm rad}$ is presented in next Fig.~\ref{fig.4},
which decrease in the region of tunneling.
\begin{figure}[h]
\centerline{
\includegraphics[width=55mm]{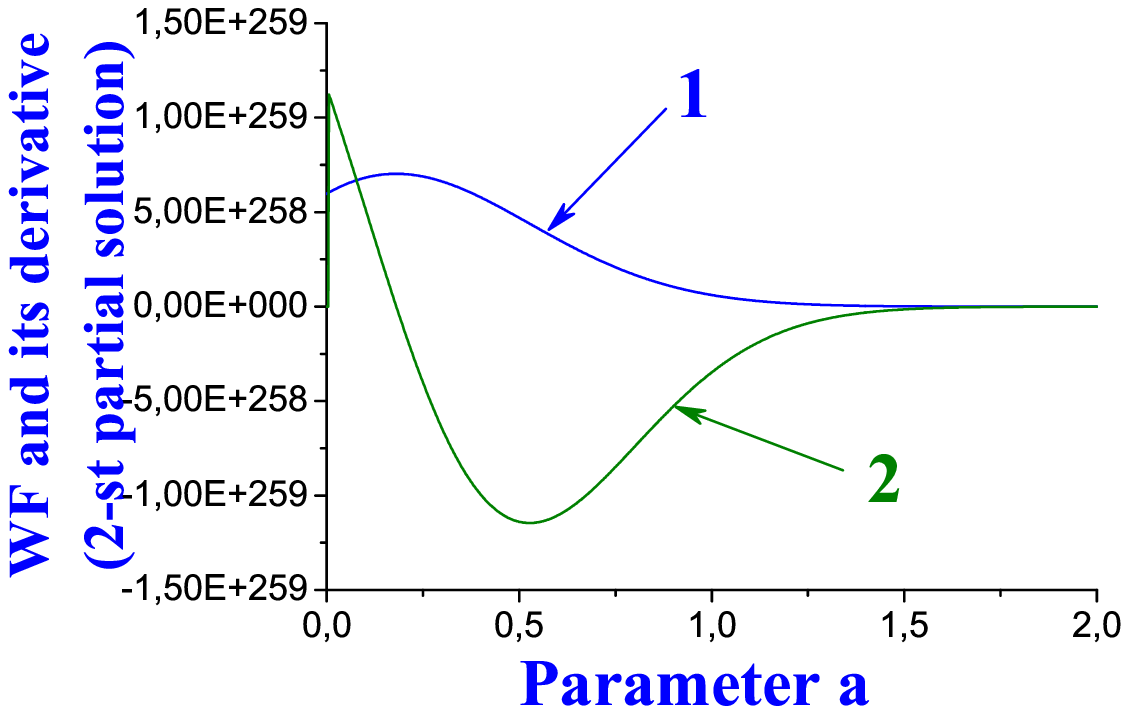}
\includegraphics[width=55mm]{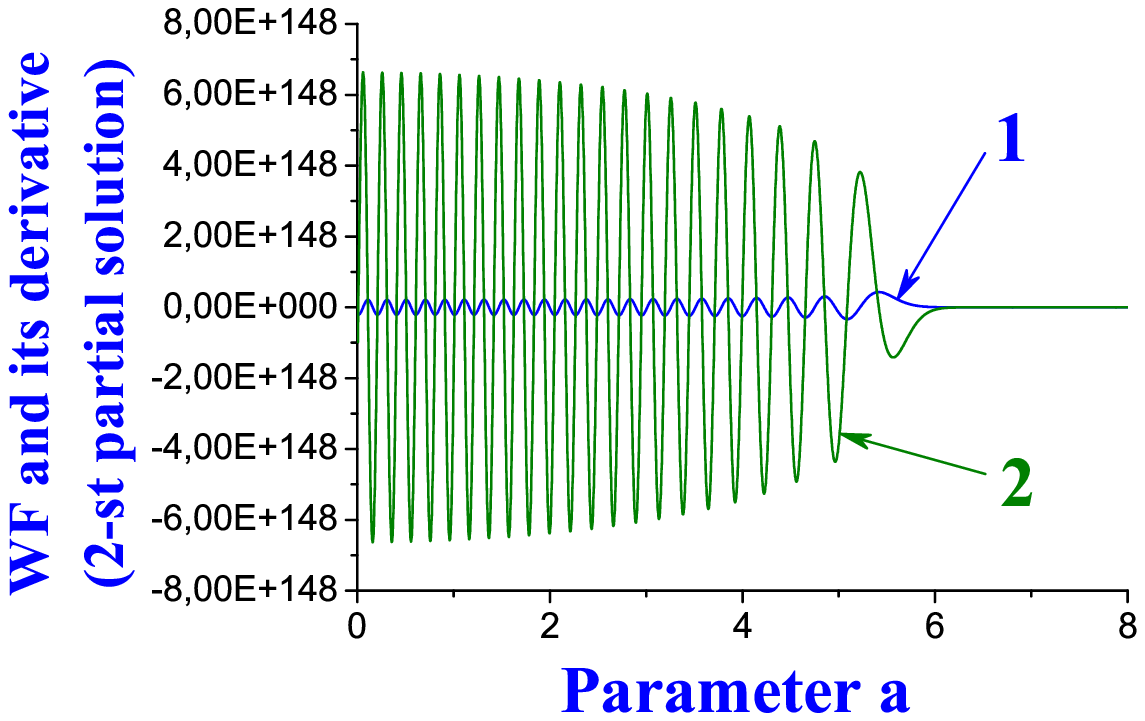}
\includegraphics[width=55mm]{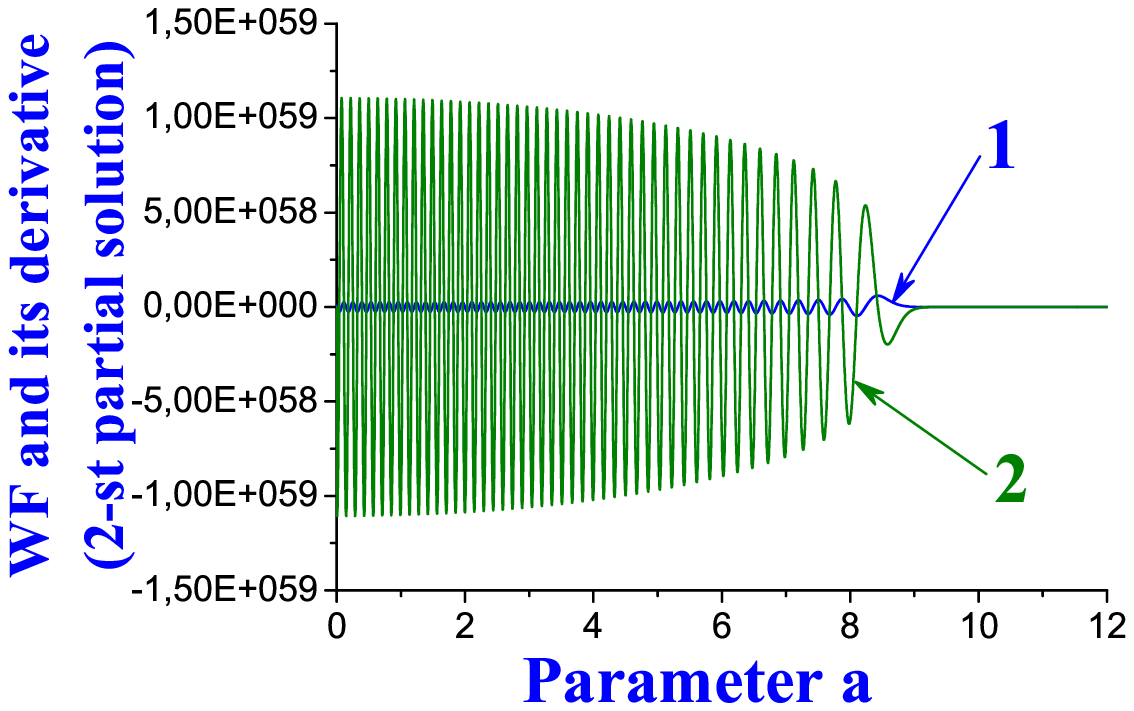}}
\caption{\small The second partial solution for the wave function and its derivative at different values of the energy of radiation $E_{\rm rad}$, decreasing in the tunneling region (curve 1, blue, is for the wave function; curve 2, green, for the derivative of this wave function):
(a) $E_{\rm rad}=10$; (b) $E_{\rm rad}=1000$; (c) $E_{\rm rad}=2000$
\label{fig.4}}
\end{figure}
According to analysis, this solution close to the turning points, in the tunneling region, in the sub-barrier and above-barrier regions looks like the first partial solution, but with a difference that now maximums of the wave function and its derivative are larger essentially in the external region in a comparison with the internal region, and amplitudes in the tunneling region are decreased monotonously.

Comparing all these pictures of the wave function with the results of Ref.~\cite{AcacioDeBarros.2007.PRD}, one can see that the wave function in such an approach is essentially more continuous, has no any divergencies and its behavior is clear anywhere. From here I conclude that \emph{in determination of the wave function and its derivative at arbitrary energy of radiation the developed method is essentially more quick, stable and accurate in a comparison with the non-stationary quantum approach in Ref.~\cite{AcacioDeBarros.2007.PRD}}.
Note the following.
\begin{itemize}

\item
With increasing $a$, a period of the oscillations both for the wave function and its derivative decreases uniformly in the external region and increases uniformly in the internal region (it partially obtained earlier in Ref.~\cite{Maydanyuk.2008.EPJC} at $E_{\rm rad}=0$).

\item
At larger distance from the barrier (i.~e. at increasing of $a$ in the external region and at decreasing of $a$ in the internal region) it becomes \underline{more difficult} to achieve the convergent continuous solutions for the wave function and its derivative (it partially obtained earlier in Ref.~\cite{Maydanyuk.2008.EPJC} at $E_{\rm rad}=0$).

\item
\emph{A number of oscillations of the wave function in the internal region increases with increasing of the energy of radiation $E_{\rm rad}$} (a new opened property).
\end{itemize}

\section{A fully quantum definition of ``free'' wave inside strong fields
\label{sec.5}}

\subsection{A problem of definition of ``free'' wave in cosmology and correction of the boundary condition
\label{sec.5.1}}

Which boundary condition should be used for such a wave function, which describes leaving of the wave from the barrier outside the most correctly and accurately? A little variation of the boundary condition leads to change of the fluxes concerning the barrier and, as result, it changes the coefficients of penetrability and reflection. So, a proper choice of the boundary condition is extremely important. However before, let us analyze how much the choice of the boundary condition
is natural in the asymptotic region.
\begin{itemize}
\item
In tasks of decay in nuclear and atomic physics such potentials of interactions are used, which tend to zero in the asymptotic region. In these tasks an application of the boundary condition at limit of infinity does not give questions. In cosmology we deal with another, principally different type of the potential: with increasing of the scale factor $a$ modulus of this potential increases. A gradient of the potential, used with opposite sign and having a sense of force acting on the wave, increases also. Therefore, \underline{here there is nothing mutual with free propagation of the wave in the asymptotic region}. \emph{Thus, a direct passage of the application of the boundary condition in the asymptotic region into cosmological problems looks very questionable.}

\item
Results in Ref.~\cite{Maydanyuk.2008.EPJC} reinforce a seriousness of this problem, which show that with increasing of the scale factor $a$ the region is enlarged where solutions for the wave function (and its two partial solutions) are stable. According to~Ref.~\cite{Maydanyuk.2008.EPJC}, the scale factor $a$ in the external region is larger, the a period of oscillations of each partial solution for the wave function is \underline{smaller}. This requires with increasing of $a$ to decrease continuously step of calculations and to increase time of calculations of the wave function for keeping the same accuracy. This increases errors in calculation of the wave (if to set it to be free in the asymptotic region), and in result the final found solution can have nothing mutual with proper one! From here a natural conclusion follows on a principal impossibility to use practically the boundary condition at infinity for calculation of the wave (in supposition if we know it maximally accurately in the asymptotic region), if we like to pass from the semiclassical approach to the fully quantum one. Principally another situation exists in problems of decay in nuclear and atomic physics where calculations of the wave in the asymptotic region are the most stable and accurate!

\item
One can add a fact that it has not been known yet whether the Universe expands at extremely large value of the scale factor $a$. Just the contrary, it would like to clarify this from a solution of the problem, however imposing a condition that the Universe expands in the initial stage.
\end{itemize}
On such a basis, I shall introduce the following \underline{\textbf{definition of the boundary condition:}}
\begin{quote}
\emph{The boundary condition should fix the wave function so that it represents the wave, interaction between which and the potential barrier is minimal at such a value of the scale factor $a$ where action of this potential is minimal as possible.}
\end{quote}
A propagation of the wave defined by such a way will be closed maximally to free one for the used potential and at used value of the scale factor $a$ (I call such a wave conditionally ``free''). However, with a purpose to give a mathematical formulation for this definition we are confronted with two questions, which we will have to resolve:

\begin{enumerate}
\item
What should the free wave represent in general in a field of the cosmological potential of arbitrary shape? How could it be defined the most correctly in enough small neighborhood of arbitrary selected coordinate?

\item
At which coordinate is the application of such a boundary condition the most corrected?
\end{enumerate}

At first, let us consider the second question where it needs to impose the boundary condition on the wave function.
Which should this point be: or this is a turning point (where the potential coincides with energy of radiation), or this is a coordinate where a gradient from the potential (having a sense of \emph{force of interaction}) becomes zero, or this is a coordinate where the potential becomes zero? With a purpose to define the wave with free propagation, we are looking for such a coordinate where the action by the potential on this wave is
minimal.
\begin{quote}
\emph{We define this coordinate where the force acting on the wave is minimal. Here, we define the force as the gradient of the potential used with opposite sign.}
\end{quote}
It turns out that according to such a (local) definition the force in the external region is minimal at the external turning point $a_{\rm tp,\,out}$. Also, the force, minimally acting on the wave incident on the barrier in the internal region and on the wave reflected from it, is minimal at the internal turning point $a_{\rm tp,\,in}$. Thus, we have just obtained the internal and external turning points where we should impose the boundary conditions in determination of the waves.
At zero energy null of the potential coincides with the turning point and therefore a choice of the point of leaving of the wave from the barrier outside in imposing of the boundary condition in Ref.~\cite{Maydanyuk.2008.EPJC} was  made properly. Here, it could be interesting to analyze a condition of equality to zero of the gradient of the potential. But this point is located in the region of the barrier where there is tunneling and, therefore, we shall not study this case in this paper.


\subsection{Definition of the wave minimally interacting with the potential
\label{sec.5.2}}

Now we shall be looking for a form of the wave function in the external region, which describes maximally accurately the wave, propagation of which is the closest maximally to ``free'' one in the external region at the turning point $a_{\rm tp,\, out}$ and is directed outside. Let us return back to eq.~(\ref{eq.model.6.6}) where the variable $q = a - a_{\rm tp,\, out}$ has been introduced.
Changing this variable to
\begin{equation}
  \xi = \bigl|V_{1}^{\rm (out)}\bigr|^{1/3} q,
\label{eq.5.2.1}
\end{equation}
this equation is transformed into
\begin{equation}
  \displaystyle\frac{d^{2}}{d\xi^{2}}\, \varphi(\xi) + \xi \: \varphi(\xi) = 0.
\label{eq.5.2.2}
\end{equation}

From quantum mechanics we know two linearly independent exact solutions for the function $\varphi(\xi)$ in this equation --- these are the \emph{Airy functions} ${\rm Ai}\,(\xi)$ and ${\rm Bi}\,(\xi)$ (for example, see Ref.~\cite{Abramowitz.1964}, p.~264--272, 291--294). Expansions of these functions into power series at small $\xi$, their asymptotic expansions at large $|\xi|$, their representations through Bessel functions, zeroes and their asymptotic expansions are known. We have some integrals of these functions, and also the form of the Airy functions in the semiclassical approximation (which can be applied at large $|\xi|$).
In some problems of the analysis of finite solutions $\varphi(\xi)$ in the whole range of $\xi$ it is convenient to use the integral representations of the Airy functions (see eq.~(10.4.32) in Ref.~\cite{Abramowitz.1964}, p.~265, $a=1/3$, taking into account sign in eq.~(10.4.1)):
\begin{equation}
\begin{array}{ccl}
  {\rm Ai} \, (\pm\xi) & = &
  \displaystyle\frac{1}{\pi} \displaystyle\int\limits_{0}^{+\infty}
  \cos{\biggl(\displaystyle\frac{u^{3}}{3} \mp \xi u \biggr)} \; du, \\

  {\rm Bi}\, (\pm\xi) & = &
  \displaystyle\frac{1}{\pi} \displaystyle\int\limits_{0}^{+\infty}
  \biggl[
    \exp{\biggl(-\displaystyle\frac{u^{3}}{3} \mp\xi u \biggr)} +
    \sin{\biggl(\displaystyle\frac{u^{3}}{3} \mp\xi u \biggr)}
  \biggr] \; du.
\end{array}
\label{eq.5.2.3}
\end{equation}
Furthermore, we shall be interested in the solution of the function $\varphi(\xi)$ which describes the most accurately the \emph{outgoing wave} in the range of $a$ close to the $a_{tp}$ point. However, it has been not clear what the wave represents in general near the point $a_{tp}$ in the potential studied, and which linear combination of the ${\rm Ai}\,(\xi)$ and ${\rm Bi}\,(\xi)$ functions defines it in the most accurate way.

The clearest and most natural understanding of the outgoing wave is given by the semiclassical consideration of the tunneling process. However, at the given potential the semiclassical approach allows us to define the outgoing wave in the asymptotic region only (while we can join solutions in the proximity of $a_{tp}$ by the Airy functions). But it is not clear whether the wave, defined in the asymptotic region, remains outgoing near the $a_{tp}$. During the whole path of its propagation outside the barrier the wave interacts with the potential, which must inevitably lead to a deformation of its shape (like to appearance of a phase shift in the scattering of a wave by a radial potential caused by interaction in scattering theory). Whether does it turn out the potentials used in cosmological models give a \underline{significantly larger change of the shape of the wave caused by interaction} in a comparison with the potentials used, for example, for the description of nuclear collisions in the framework of scattering theory? Moreover, for the given potential there is a problem with obtaining convergence in the calculation of the partial solutions for the wave function in the asymptotic region. According to our calculations, a small change in the range of the definition of the wave in the asymptotic region leads to a significant increase of errors, which requires one to increase the accuracy of the calculations.
Therefore, we shall be looking for a way of defining the outgoing wave not in the asymptotic region, but in the closest vicinity of the point of escape, $a_{tp}$. In a search of solutions close to the point $a_{tp}$, i.~e. at small enough $|\xi|$, the validity of the semiclassical method breaks down as $|\xi|$ approaches zero. Therefore, we shall not use the semiclassical approach in this paper.

Assuming the potential $V(a)$ to have an arbitrary form, we define the wave at the point $a_{tp}$ in the following way.
\begin{dfn}[strict definition of the wave]
\label{def.wave.strict}
The wave is such a linear combination of two partial solutions of the wave function that the change of the modulus $\rho$ of this wave function is closest to constant under variation of $a$:
\begin{equation}
  \displaystyle\frac{d^{2}}{da^{2}}\, \rho(a) \biggl|_{a=a_{tp}} \to 0.
\label{eq.5.2.4}
\end{equation}
\end{dfn}
According to this definition, the real and imaginary parts of the total wave function have the mutually closest behaviors under the same variation of $a$, and the difference between possible maximums and minimums of the modulus of the total wave function is the smallest. For some types of potentials (in particular, for a rectangular barrier) it is more convenient to define the wave less strongly.

\vspace{3mm}
\noindent
\underline{\textbf{Definition 2 (weak definition of wave):}}
\begin{quote}
\emph{The wave is such a linear combination of two partial solutions of wave function that the modulus $\rho$ changes minimally under variation of $a$:}
\begin{equation}
  \displaystyle\frac{d}{da}\, \rho(a) \biggl|_{a=a_{tp}} \to 0.
\label{eq.5.2.5}
\end{equation}
\end{quote}
According to this definition, the change of the wave function caused by variation of $a$ is characterized mainly by its phase (which can characterize the interaction between the wave and the potential).

Subject to this requirement, we shall look for the solution for the function $\varphi(\xi)$ in the following form:
\begin{equation}
  \varphi\, (\xi) = T \cdot \Psi^{(+)}(\xi),
\label{eq.5.2.6}
\end{equation}
where
\begin{equation}
\begin{array}{ccl}
  \Psi^{(\pm)} (\xi) & = &
    \displaystyle\int\limits_{0}^{u_{\rm max}}
    \exp{\pm\,i\,\Bigl(-\displaystyle\frac{u^{3}}{3} + f(\xi)\,u \Bigr)} \; du.
\end{array}
\label{eq.5.2.7}
\end{equation}
where $T$ is an unknown normalization factor, $f(\xi)$ is an unknown continuous function satisfying $f(\xi) \to {\rm const}$ at $\xi \to 0$, and $u_{\rm max}$ is the unknown upper limit of integration. In such a solution, the real part of the function $f(\xi)$ gives a contribution to the phase of the integrand function only while the imaginary part of $f(\xi)$ deforms its modulus.

Let us find the first and second derivatives of the function $\Psi(\xi)$ (a prime denotes a derivative with respect to $\xi$):
\begin{equation}
\begin{array}{ccl}
  \displaystyle\frac{d}{d\xi}\, \Psi^{(\pm)} (\xi) & = &
    \pm\, i \displaystyle\int\limits_{0}^{u_{\rm max}} f^{\prime} u \;
    \exp{\pm i\, \Bigl(-\displaystyle\frac{u^{3}}{3} + f(\xi) u \Bigr)} \; du, \\
  \displaystyle\frac{d^{2}}{d\xi^{2}}\, \Psi^{(\pm)} (\xi) & = &
    \displaystyle\int\limits_{0}^{u_{\rm max}}
    \Bigl(\pm\, if^{\prime\prime} u - (f^{\prime})^{2} u^{2} \Bigr) \:
    \exp{\pm i\, \Bigl(-\displaystyle\frac{u^{3}}{3} + f(\xi) u \Bigr)} \; du.
\end{array}
\label{eq.5.2.8}
\end{equation}
From this we obtain:
\begin{equation}
\begin{array}{c}
  \displaystyle\frac{d^{2}}{d\xi^{2}}\, \Psi^{(\pm)} (\xi) + \xi\: \Psi^{(\pm)} (\xi) =
  \displaystyle\int\limits_{0}^{u_{\rm max}}
    \Bigl(\pm\, if^{\prime\prime} u - (f^{\prime})^{2} u^{2} + \xi \Bigr) \:
    \exp{\pm i\, \Bigl(-\displaystyle\frac{u^{3}}{3} + f(\xi) u \Bigr)} \; du.
\end{array}
\label{eq.5.2.9}
\end{equation}

Considering the solutions at small enough values of $|\xi|$, we represent $f(\xi)$ in the form of a power series:
\begin{equation}
  f(\xi) = \sum\limits_{n=0}^{+\infty} f_{n}\, \xi^{n},
\label{eq.5.2.10}
\end{equation}
where $f_{n}$ are constant coefficients. The first and second derivatives of $f(\xi)$ are
\begin{equation}
\begin{array}{l}
  f^{\prime}(\xi) = \displaystyle\frac{d}{d\xi}\, f(\xi) =
    \sum\limits_{n=1}^{+\infty} n f_{n} \; \xi^{n-1} = \sum\limits_{n=0}^{+\infty} (n+1) \: f_{n+1} \; \xi^{n}, \\
  f^{\prime\prime}(\xi) = \displaystyle\frac{d^{2}}{d\xi^{2}}\, f(\xi) =
    \sum\limits_{n=0}^{+\infty} (n+1)\,(n+2) \: f_{n+2} \; \xi^{n}.
\end{array}
\label{eq.5.2.11}
\end{equation}
Substituting these solutions into eq.~(\ref{eq.5.2.8}), we obtain
\begin{equation}
\begin{array}{c}
  \displaystyle\frac{d^{2}}{d\xi^{2}}\, \Psi^{(\pm)} (\xi) + \xi\: \Psi^{(\pm)} (\xi) =
  \displaystyle\int\limits_{0}^{u_{\rm max}}
    \Biggl\{
      \Bigl(\pm\,2iu \: f_{2} - u^{2} \: f_{1}^{2} \Bigr) +
      \Bigl(\pm\,6iu \: f_{3} - 4 u^{2} \: f_{1}f_{2} + 1 \Bigr) \: \xi + \\
    + \sum\limits_{n=2}^{+\infty} \Bigl[\pm\,iu \: (n+1)(n+2) \: f_{n+2} -
      u^{2} \sum\limits_{m=0}^{n} (n-m+1)(m+1) \: f_{n-m+1}f_{m+1} \Bigr] \: \xi^{n}
    \Biggr\} \:
    \exp{\pm\,i \Bigl(-\displaystyle\frac{u^{3}}{3} + fu \Bigr)} \; du.
\end{array}
\label{eq.5.2.12}
\end{equation}

Considering this expression at small $|\xi|$, we use the following approximation:
\begin{equation}
\begin{array}{ccc}
  \exp{\pm\,i \Bigl(-\displaystyle\frac{u^{3}}{3} + fu \Bigr)} & \to &
  \exp{\pm\,i \Bigl(-\displaystyle\frac{u^{3}}{3} + f_{0}u \Bigr)}.
\end{array}
\label{eq.5.2.13}
\end{equation}
Then from eq.~(\ref{eq.5.2.2}) we obtain the following condition for the unknown $f_{n}$:
\begin{equation}
\begin{array}{c}
  \displaystyle\int\limits_{0}^{u_{\rm max}}
    \Bigl(\pm\,2iu \: f_{2} - u^{2} \: f_{1}^{2} \Bigr)
    \exp{\pm i\, \Bigl(-\displaystyle\frac{u^{3}}{3} + f_{0}u \Bigr)} \; du \; + \\

  + \;
    \xi \cdot \displaystyle\int\limits_{0}^{u_{\rm max}}
    \Bigl(\pm\,6iu \: f_{3} - 4 u^{2} \: f_{1}f_{2} + 1 \Bigr) \:
    \exp{\pm i\, \Bigl(-\displaystyle\frac{u^{3}}{3} + f_{0}u \Bigr)} \; du \; + \\

  + \;
    \sum\limits_{n=2}^{+\infty}  \xi^{n} \cdot
    \displaystyle\int\limits_{0}^{u_{\rm max}}
    \Bigl[\pm\, iu \: (n+1)(n+2) \: f_{n+2} -
    u^{2} \sum\limits_{m=0}^{n} (n-m+1)(m+1) \: f_{n-m+1}f_{m+1} \Bigr] \:
    \exp{\pm i\, \Bigl(-\displaystyle\frac{u^{3}}{3} + f_{0}u \Bigr)} \; du \; = 0.
\end{array}
\label{eq.5.2.14}
\end{equation}
Requiring this condition to be satisfied for different $\xi$ with different powers $n$, we obtain the following system:
\begin{equation}
\begin{array}{cc}
\xi^{0}: &
  \displaystyle\int\limits_{0}^{u_{\rm max}}
    \Bigl(\pm\,2iu \: f_{2} - u^{2} \: f_{1}^{2} \Bigr) \:
    \exp{\pm i\, \Bigl(-\displaystyle\frac{u^{3}}{3} + f_{0}u \Bigr)} \; du = 0, \\

\xi^{1}: &
  \displaystyle\int\limits_{0}^{u_{\rm max}}
      \Bigl(\pm\, 6iu \: f_{3} - 4 u^{2} \: f_{1}f_{2} + 1 \Bigr) \:
    \exp{\pm i\, \Bigl(-\displaystyle\frac{u^{3}}{3} + f_{0}u \Bigr)} \; du = 0, \\

\xi^{n}: &
  \displaystyle\int\limits_{0}^{u_{\rm max}}
    \Bigl[\pm\, iu \: (n+1)(n+2) \: f_{n+2} -
    u^{2} \sum\limits_{m=0}^{n} (n-m+1)(m+1) \: f_{n-m+1}f_{m+1} \Bigr] \:
    \exp{\pm i\, \Bigl(-\displaystyle\frac{u^{3}}{3} + f_{0}u \Bigr)} \; du = 0.
\end{array}
\label{eq.5.2.15}
\end{equation}

Assuming the coefficients $f_{0}$ and $f_{1}$ to be given, we find the following solutions for the unknown $f_{2}$, $f_{3}$ and $f_{n}$:
\begin{equation}
\begin{array}{ccc}
  f_{2}^{(\pm)} = \pm\;\displaystyle\frac{f_{1}^{2}}{2i} \cdot \displaystyle\frac{J_{2}^{(\pm)}}{J_{1}^{(\pm)}}, &
  f_{3}^{(\pm)} =
    \pm\;\displaystyle\frac{4 f_{1}f_{2}^{(\pm)}\, J_{2}^{(\pm)} - J_{0}^{(\pm)}} {6i\, J_{1}^{(\pm)}}, &
  f_{n+2}^{(\pm)} =
    \displaystyle\frac{\sum\limits_{m=0}^{n} (n-m+1)(m+1) \: f_{n-m+1}^{(\pm)}\,f_{m+1}^{(\pm)}}
      {i \: (n+1)(n+2)} \cdot
    \displaystyle\frac{J_{2}^{(\pm)}} {J_{1}^{(\pm)}},
\end{array}
\label{eq.5.2.16}
\end{equation}
where the following notations for the integrals have been introduced:
\begin{equation}
\begin{array}{ccc}
  J_{0}^{(\pm)} =
    \displaystyle\int\limits_{0}^{u_{\rm max}}
    \exp{\pm i\, \Bigl(-\displaystyle\frac{u^{3}}{3} + f_{0}u \Bigr)} \; du, &
  J_{1}^{(\pm)} =
    \displaystyle\int\limits_{0}^{u_{\rm max}}
    u \: \exp{\pm i\, \Bigl(-\displaystyle\frac{u^{3}}{3} + f_{0}u \Bigr)} \; du, &
  J_{2}^{(\pm)} =
    \displaystyle\int\limits_{0}^{u_{\rm max}}
    u^{2} \: e^{\pm i\, \Bigl(-\displaystyle\frac{u^{3}}{3} + f_{0}u \Bigr)} \; du.
\end{array}
\label{eq.5.2.17}
\end{equation}
Thus, we see that the solution (\ref{eq.5.2.6}) with taking into account eq.~(\ref{eq.5.2.7}) for the function $\varphi\,(\xi)$ has arbitrariness in a choice of the unknown coefficients $f_{0}$, $f_{1}$ and the upper limit of integration $u_{\rm max}$. However, the solutions found, eqs.~(\ref{eq.5.2.16}), define the function $f(\xi)$ so as to ensure that the equality (\ref{eq.5.2.6}) is \underline{exactly} satisfied in the region of $a$ close to the escape point $a_{tp}$.
This proves that \emph{the function $\varphi\,(\xi)$ in the form (\ref{eq.5.2.6}) with taking into account eq.~(\ref{eq.5.2.7}) at an arbitrary choice of $f_{0}$, $f_{1}$ and $u_{\rm max}$ is the \underline{exact} solution of the Schr\"{o}dinger equation near the escape point $a_{tp}$}. In order to bring the solution $\Psi(\xi)$ into the well-known Airy functions, ${\rm Ai}\,(\xi)$ and ${\rm Bi}\,(\xi)$,
we select
\begin{equation}
\begin{array}{cc}
  f_{0} = 0, &
  f_{1} = 1.
\end{array}
\label{eq.5.2.18}
\end{equation}
At such a choice of the coefficients $f_{0}$ and $f_{1}$, the integrand function in the solution (\ref{eq.5.2.7}) up to $\xi^{2}$ has a constant modulus and a varying phase (the coefficient $f_{2}$ deforms the modulus, but it is fulfilled at $\xi^{3}$). Therefore, one can expect that the solution (\ref{eq.5.2.6}) at the turning point $a_{tp}$ describes the wave with a maximally proper shape.


\subsection{Total wave function
\label{sec.5.3}}

\begin{figure}[h]
\centerline{
\includegraphics[width=55mm]{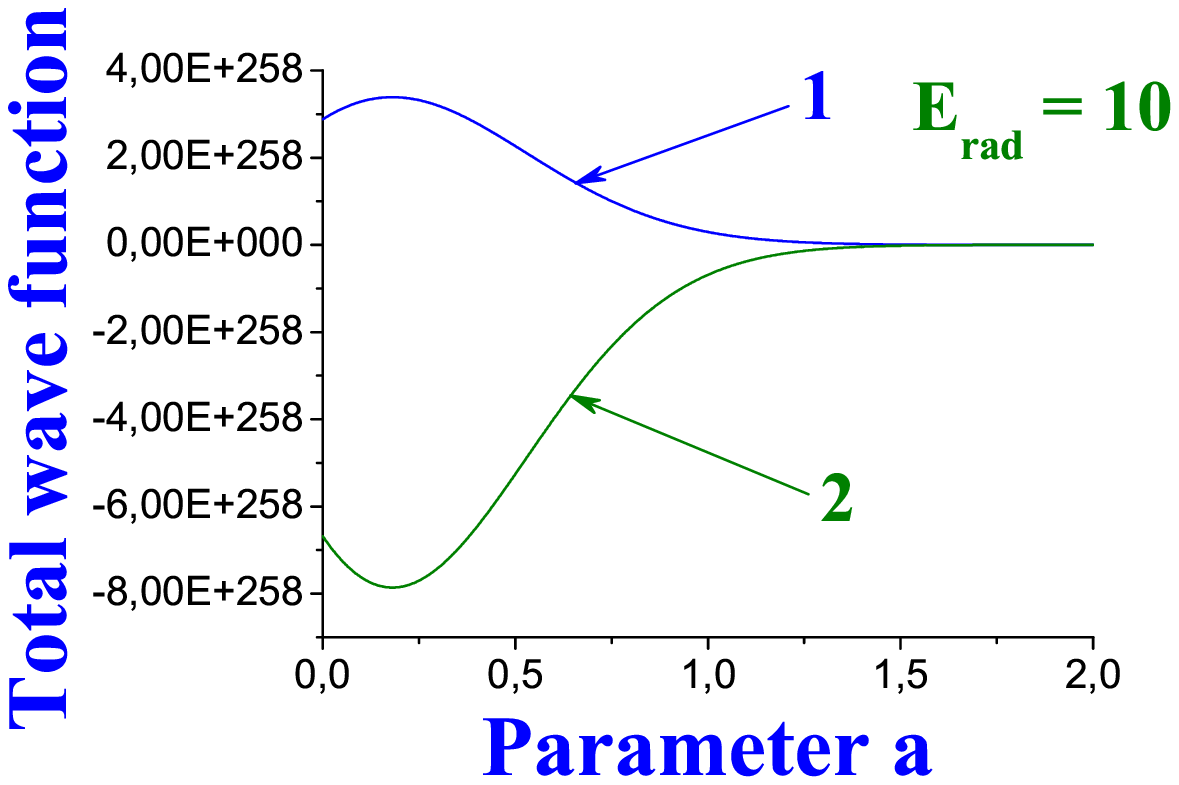}
\includegraphics[width=55mm]{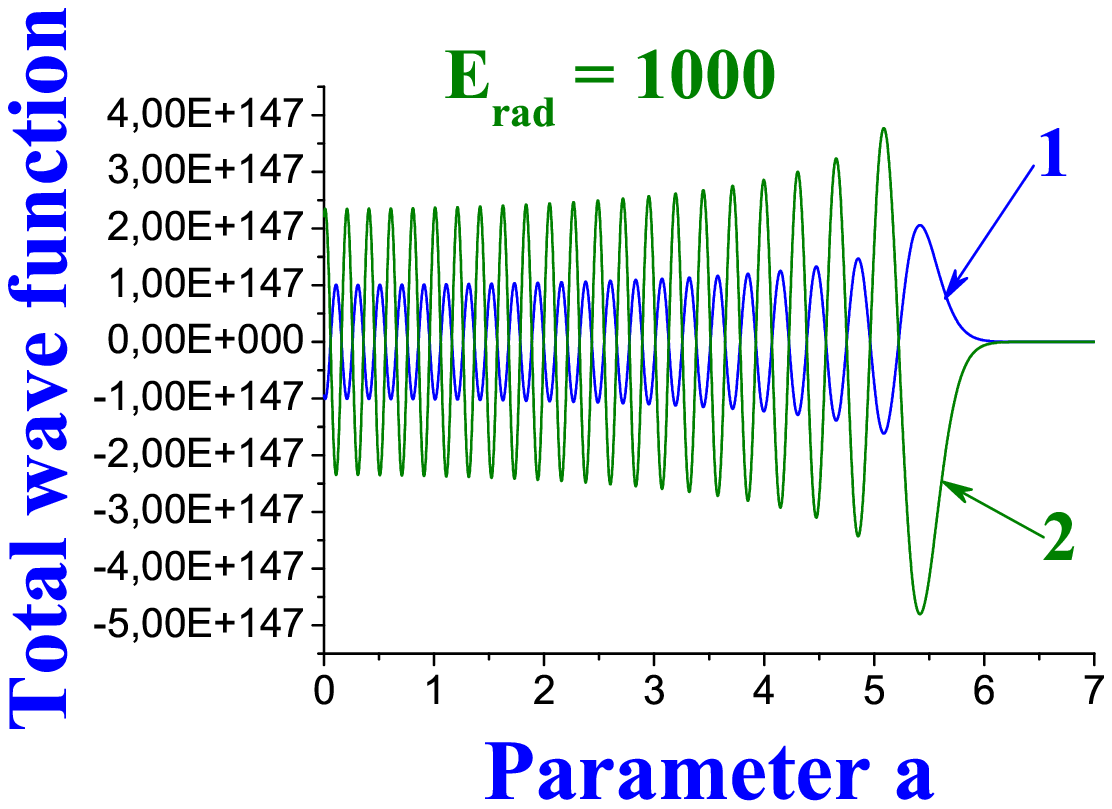}
\includegraphics[width=55mm]{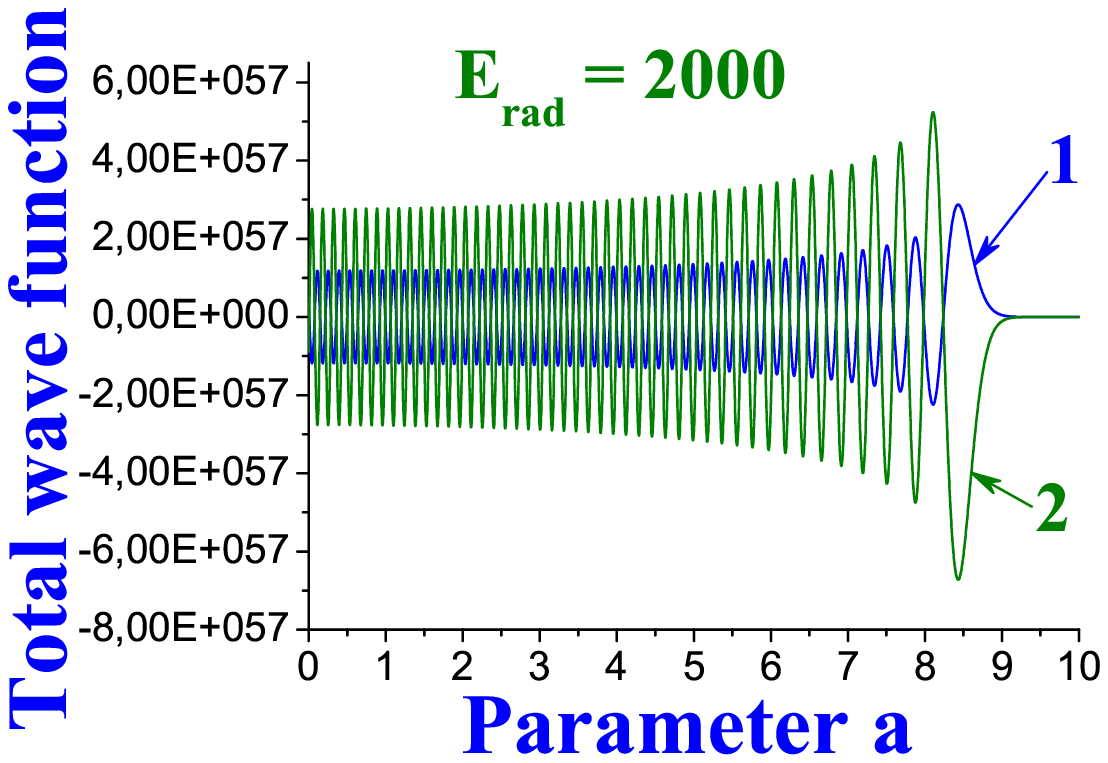}}
\caption{\small The wave function at selected values of the energy of radiation $E_{\rm rad}$ (curve 1, blue, is for real part of the wave function; curve 2, green, for imaginary part of the wave function):
(a) $E_{\rm rad}=10$; (b) $E_{\rm rad}=1000$; (c) $E_{\rm rad}=2000$
\label{fig.5}}
\end{figure}

Having obtained two linearly independent partial solutions $\varphi_{1}(a)$ and $\varphi_{2}(a)$,
we make up a general solution (a prime is for the derivative with respect to $a$):
\begin{equation}
  \varphi\,(a) = T \cdot \bigl(C_{1}\, \varphi_{1}(a) + C_{2}\,\varphi_{2}(a) \bigr),
\label{eq.5.3.1}
\end{equation}
\begin{equation}
\begin{array}{cc}
  C_{1} = \displaystyle\frac{\Psi\varphi_{2}^{\prime} - \Psi^{\prime}\varphi_{2}}
          {\varphi_{1}\varphi_{2}^{\prime} - \varphi_{1}^{\prime}\varphi_{2}} \bigg|_{a=a_{\rm tp,\, out}}, &
  C_{2} = \displaystyle\frac{\Psi^{\prime}\varphi_{1} - \Psi\varphi_{1}^{\prime}}
          {\varphi_{1}\varphi_{2}^{\prime} - \varphi_{1}^{\prime}\varphi_{2}} \bigg|_{a=a_{\rm tp,\, out}},
\end{array}
\label{eq.5.3.3}
\end{equation}
where $T$ is normalization factor, $C_{1}$ and $C_{2}$ are complex constants found from the boundary condition introduced above: \emph{the $\varphi\,(a)$ function should represent an outgoing wave at turning point $a_{\rm tp,\, out}$}.
%
%
The total wave function calculated by such a way for the potential (\ref{eq.model.5.9}) with parameters $A=36$, $B=12\,\Lambda$ at $\Lambda=0.01$
at different values of the energy of radiation $E_{\rm rad}$ is shown in Fig.~\ref{fig.5}.
One can see that a number of oscillations of the wave function in the internal region increases with increasing of the energy of radiation. Another interesting property is \emph{the larger maximums of the wave function in the internal region at the smaller distances to the barrier for arbitrary energy} (it has been found for the first time).

In the next Fig.~\ref{fig.6} it has shown how a modulus of this wave function is changed at selected values of the energy of radiation.
\begin{figure}[h]
\centerline{
\includegraphics[width=55mm]{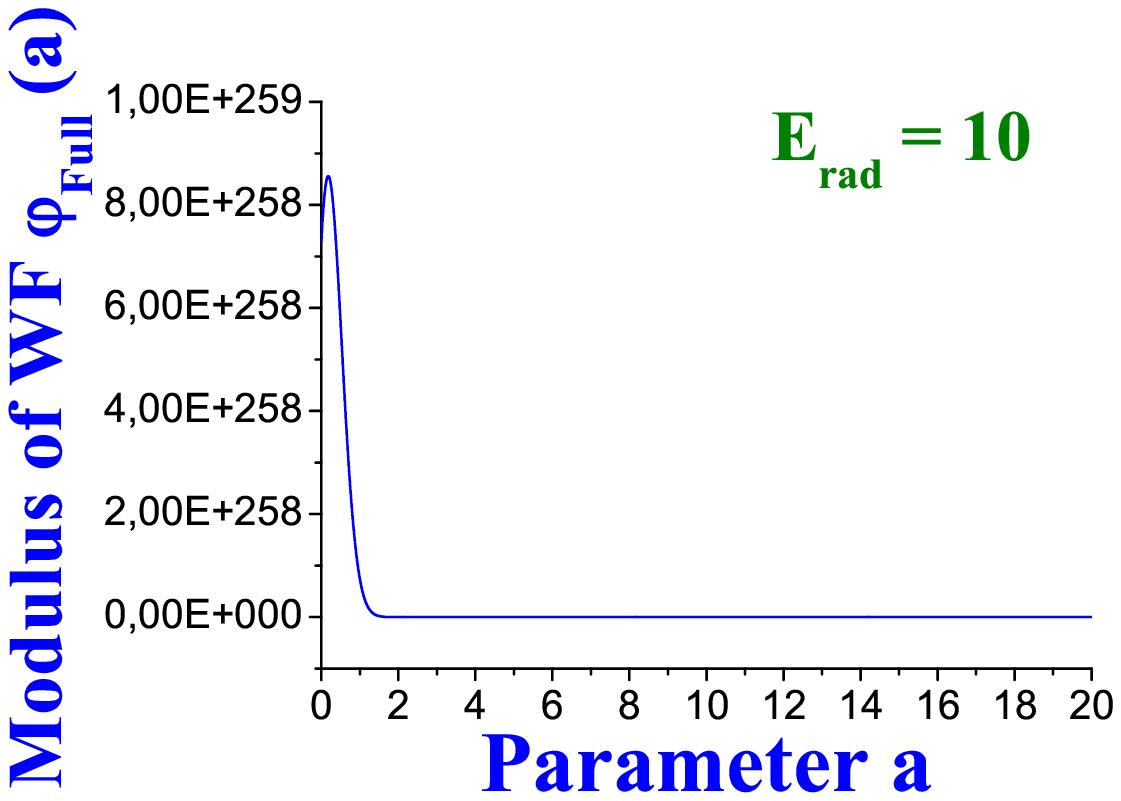}
\includegraphics[width=55mm]{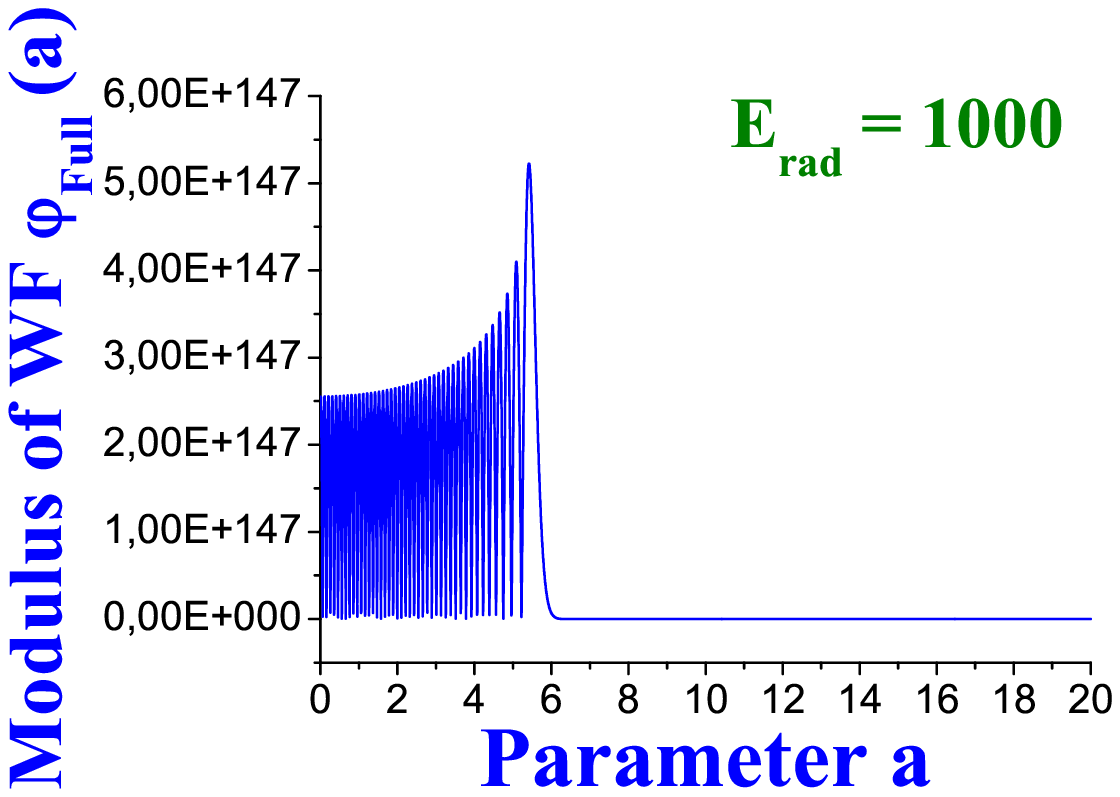}
\includegraphics[width=55mm]{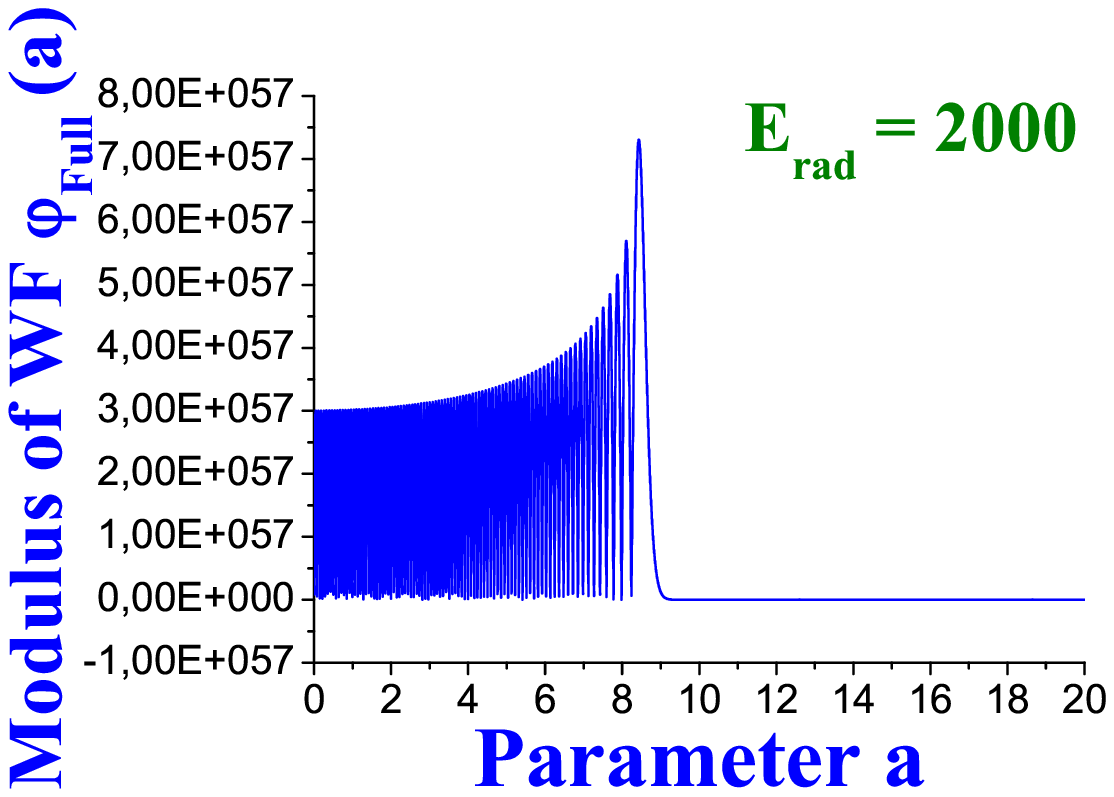}}
\caption{\small A behavior of the modulus of the wave function at the selected energies of radiation $E_{\rm rad}$:
(a) $E_{\rm rad}=10$; (b) $E_{\rm rad}=1000$; (c) $E_{\rm rad}=2000$.
\label{fig.6}}
\end{figure}
From these figures it becomes clear why the coefficient of penetrability of the barrier is extremely small (up to the energy $E_{\rm rad}=2000$). In order to estimate, how much the boundary condition introduced above is effective in construction of the wave on the basis of the total wave function close to the external turning point $a_{\rm tp,\,out}$, it is useful to see how the modulus of this wave function is changed close to this point. In Fig.~\ref{fig.7} the modulus of the found wave function close to the turning points
at the energy of radiation $E_{\rm rad}=2000$ is shown.
\begin{figure}[h]
\centerline{
\includegraphics[width=55mm]{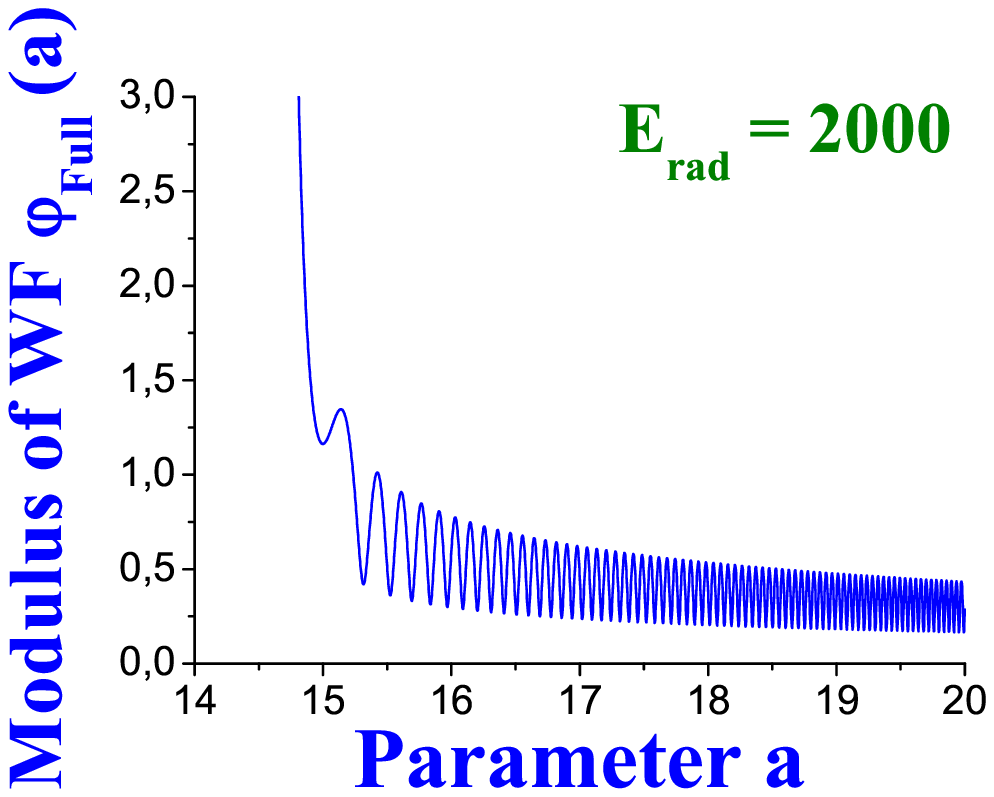}
\includegraphics[width=55mm]{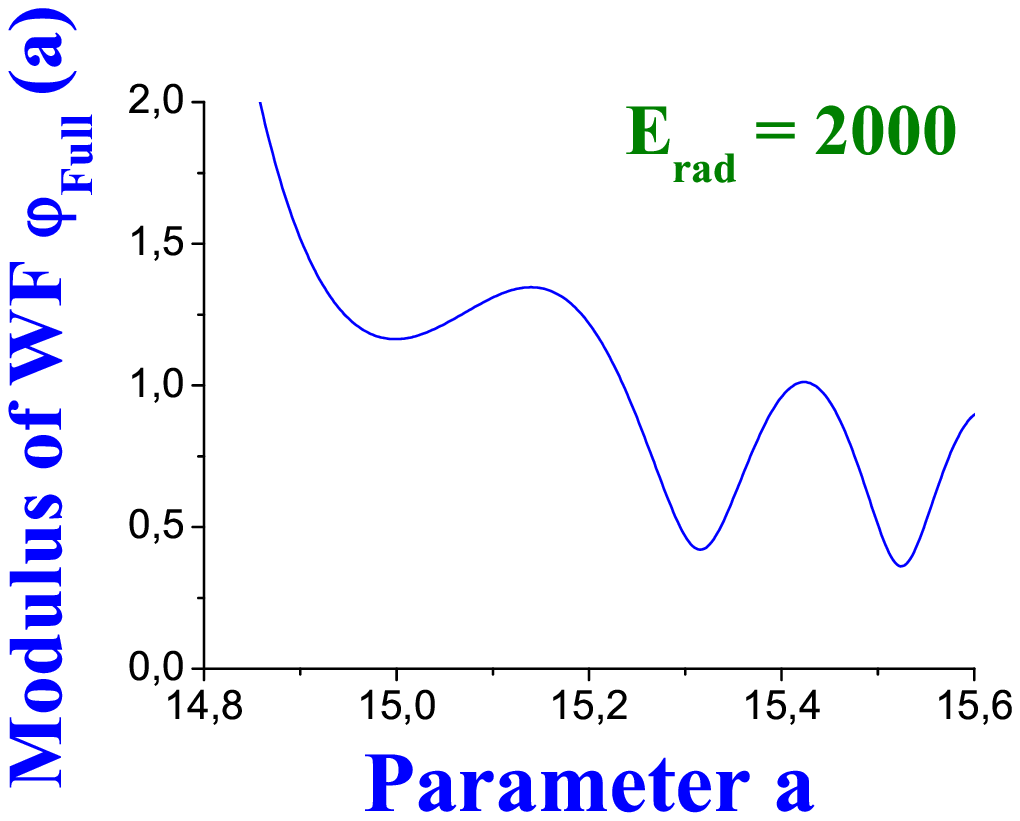}
\includegraphics[width=55mm]{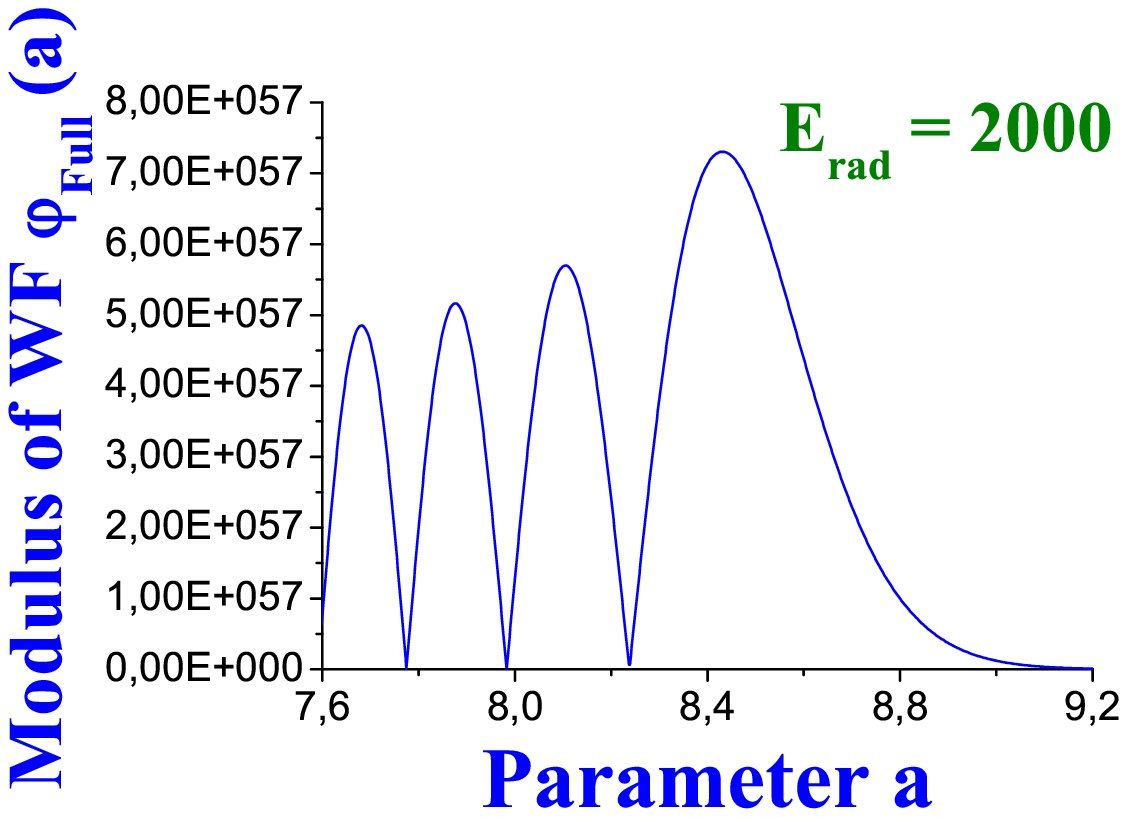}}
\caption{\small A behavior of the modulus of the total wave function at the energy of radiation $E_{\rm rad}=2000$ close to the turning points (I obtain: $a_{\rm tp,\,in}=8.58$, $a_{\rm tp,\,out}=15.04$, also see Table 1):
(a) the modulus, decreasing monotonously in the tunneling region, with further increasing of $a$ obtains maximums and holes connected with the oscillations of the wave function in the external region, but the modulus is not equal to zero (that points out existence of \underline{non-zero} flux);
(b) with increasing of $a$, close to the external turning point $a_{\rm tp,\,out}$ the modulus is changed minimally (this demonstrates practical fulfillment of the definition for the wave at such a point);
(c) transition back close to $a_{\rm tp,\,in}$ is shown, where at increasing of $a$ the modulus with maximums and holes is transformed rapidly into monotonously decreasing without maximums and holes, that is connected with transition to the region of tunneling
\label{fig.7}}
\end{figure}
Here, one can see that the modulus at $a_{\rm tp,\,out}$ is constant practically (see left panel in Fig.~\ref{fig.7}). It is interesting to note that the modulus of so defined wave function close to the internal turning point $a_{\rm tp,\,in}$ is not changed practically also and is close to maximum (see right panel in Fig.~\ref{fig.7}).


\section{A penetrability and reflection in the fully quantum approach
\label{sec.4}}

Let us analyze whether a known wave function in the whole region of its definition allows us to determine uniquely the coefficients of penetrability and reflection.

\subsection{Problem of interference between the incident and reflected waves
\label{sec.4.1}}

Rewriting the wave function $\varphi_{\rm total}$ in the internal region through a summation of incident $\varphi_{\rm inc}$ wave
and reflected $\varphi_{\rm ref}$ wave:
\begin{equation}
  \varphi_{\rm total} =
  \varphi_{\rm inc} + \varphi_{\rm ref},
\label{eq.4.1.1}
\end{equation}
we consider the total flux:
\begin{equation}
\begin{array}{ccccc}
  j\, (\varphi_{\rm total}) & = &
  i\,
  \biggl[
    \Bigl( \varphi_{\rm inc} + \varphi_{\rm ref} \Bigr)
    \nabla \Bigl( \varphi_{\rm inc}^{*} + \varphi_{\rm ref}^{*} \Bigr) -
    \mbox{h.~c.}
  \Bigr) \biggr] & = &
  j_{\rm inc} + j_{\rm ref} + j_{\rm mixed},
\end{array}
\label{eq.4.1.2}
\end{equation}
where
\begin{equation}
\begin{array}{ccl}
  j_{\rm inc} =
  i\, \Bigl(\varphi_{\rm inc} \nabla \varphi_{\rm inc}^{*} - \mbox{h.~c.}\Bigr), &

  j_{\rm ref} =
  i\, \Bigl(\varphi_{\rm ref} \nabla \varphi_{\rm ref}^{*} - \mbox{h.~c.}\Bigr), &

  j_{\rm mixed} =
  i\,
  \Bigl(
    \varphi_{\rm inc} \nabla \varphi_{\rm ref}^{*} +
    \varphi_{\rm ref} \nabla \varphi_{\rm inc}^{*} - \mbox{h.~c.}
  \Bigr).
\end{array}
\label{eq.4.1.3}
\end{equation}
The $j_{\rm mixed}$ component describes interference between the incident and reflected waves in the internal region (let us call it as \emph{mixed component of the total flux} or simply \emph{flux of mixing}).
From constancy of the total flux $j_{\rm total}$ we find flux $j_{\rm tr}$ for the wave
transmitted through the barrier, and:
\begin{equation}
\begin{array}{cc}
  j_{\rm inc} = j_{\rm tr} - j_{\rm ref} - j_{\rm mixed}, &
  j_{\rm tr} = j_{\rm total} = {\rm const}.
\end{array}
\label{eq.4.1.5}
\end{equation}
Now one can see that \emph{the mixed flux introduces ambiguity in determination of the penetrability and reflection for the same known wave function.}

\subsection{Locality of penetrability, reflection and mixing in the radial task
\label{sec.4.2}}

In quantum mechanics the coefficients of penetrability and reflection are defined concerning a potential as whole, including asymptotic regions. However, in the radial problem of quantum decay such a consideration looks to be conditional as the incident and reflected waves should be defined inside internal region from the left of the barrier. But, whether does choice of such a region where we use these waves, take influence on the penetrability and reflection? In order to formulate these coefficients, we shall include into definitions coordinates where the fluxes
are defined (denote them as $x_{\rm left}$ and $x_{\rm right}$):
\begin{equation}
\begin{array}{ccc}
  T(x_{\rm left}, x_{\rm right}) = \displaystyle\frac{j_{\rm tr}(x_{\rm right})}{j_{\rm inc}(x_{\rm left})}, &
  R(x_{\rm left}) = \displaystyle\frac{j_{\rm ref}(x_{\rm left})}{j_{\rm inc}(x_{\rm left})}, &
  M(x_{\rm left}) = \displaystyle\frac{j_{\rm mixed}(x_{\rm left})}{j_{\rm inc}(x_{\rm left})}.
\end{array}
\label{eq.4.2.1}
\end{equation}
So, the $T$ and $R$ coefficients determine a probability of transmission (or tunneling) and reflection of the wave relatively the region of the potential with arbitrary selected boundaries $x_{\rm left}$, $x_{\rm right}$.
Tending $x_{\rm right}$ to the asymptotic limit, so defined coefficients should transform into standard ones.
From eqs.~(\ref{eq.4.1.5}) and (\ref{eq.4.2.1}) we obtain
($j_{\rm tr}$ and $j_{\rm ref}$ are directed in opposite directions,
$j_{\rm inc}$ and $j_{\rm tr}$  --- in the same directions):
\begin{equation}
  |T| + |R| - M = 1.
\label{eq.4.2.3}
\end{equation}
\emph{Now we see that condition
$|T| + |R| = 1$
has sense in quantum mechanics only if there are no any interference between incident and reflected waves which we calculate},
and it is enough:
\begin{equation}
  j_{\rm mixed} = 0.
\label{eq.4.2.5}
\end{equation}
But a new question has been appeared: \emph{whether does this condition allow to separate the total wave function into the incident and reflected components uniquely?}
It turns out that the choice of the incident and reflected waves takes essential influence on the barrier penetrability, and different forms of the incident $\varphi_{\rm inc}$ and reflected $\varphi_{\rm ref}$ waves can give zero flux $j_{\rm mix}$.
If we liked to pass from rectangular internal well or some simple presentations of waves in semiclassical approach to fully quantum realistic consideration, the problem would become open and difficult. Such a situation is typical in quantum cosmology.
\emph{Whole importance of accurate definition of the wave in the quantum cosmological problem becomes clear from here for construction of the total wave function on the basis of its two partial solutions, and for the separation of the known wave function into the incident and reflected waves in the internal region from the left of the barrier.}


\subsection{Wave incident on the barrier and wave reflected from it in the internal region
\label{sec.5.4}}

One can define the incident wave to be proportional to the function $\Psi^{(+)}$ and
the reflected wave to be proportional to the function $\Psi^{(-)}$:
\begin{equation}
\begin{array}{cclccl}
  \varphi_{\rm total}\,(a) = \varphi_{\rm inc}\,(a) + \varphi_{\rm ref}\,(a), &
  \varphi_{\rm inc}\,(a) = I \cdot \Psi^{(+)}\,(a), &
  \varphi_{\rm ref}\,(a) = R \cdot \Psi^{(-)}\,(a),
\end{array}
\label{eq.5.4.1}
\end{equation}
where $I$ and $R$ are new constants found from continuity condition of the total wave function $\varphi_{\rm total}$ and its derivative 
at the internal turning point $a_{\rm tp,\, int}$:
\begin{equation}
\begin{array}{cclccl}
  I & = &
    \displaystyle\frac
      {\varphi_{\rm total}\,\Psi^{(-),\prime} - \varphi_{\rm total}^{\prime}\,\Psi^{(-)}}
      {\Psi^{(+)}\,\Psi^{(-),\prime} - \Psi^{(+),\prime}\,\Psi^{(-)}}
      \bigg|_{a=a_{\rm tp,\,int}}, &
  R & = &
    \displaystyle\frac
      {\varphi_{\rm total}^{\prime}\,\Psi^{(+)} - \varphi_{\rm total}\,\Psi^{(+),\prime}}
      {\Psi^{(+)}\,\Psi^{(-),\prime} - \Psi^{(+),\prime}\,\Psi^{(-)}}
      \bigg|_{a=a_{\rm tp,\,int}}.
\end{array}
\label{eq.5.4.2}
\end{equation}
On the basis of these solutions we obtain at the internal turning point $a_{\rm tp,\, int}$ the flux incident on the barrier, the flux reflected from it and the flux of mixing.
The flux transmitted through the barrier we calculate at the external turning point $a_{\rm tp,\, ext}$.

\subsection{Boundary condition at $a=0$: stationary approach versus non-stationary one
\label{sec.5.5}}

A choice of the proper boundary condition imposed on the wave function is directly connected with questions: could the wave function be defined at $a=0$, and which value should it be equal to at this point in such a case?

The complex total wave function is constructed on the basis of its two partial solutions which are calculated using methods in Appendix B. Such two partial solutions should be linearly independent, and possible ``measure'' of such linear independence (i.~e. nonlinear difference between these two solutions) could define accuracy of the final obtained penetrability. In particular, these two partial solutions can be real (not complex), without any decrease of accuracy in determination of the total wave function. \emph{For any desirable boundary condition imposed on the total wave function, such methods should be working.}
In order to ensure the highest linear independence between two partial solutions, I select one solution to be increasing in the region of tunneling and another one to be decreasing in this tunneling region. For the increasing partial solution as the starting point $a_{x}$ I use internal turning point $a_{\rm tp,\, in}$ at non-zero energy $E_{\rm rad}$ or equals to zero $a_{x}=0$ at null energy $E_{\rm rad}$, and for the second decreasing partial solution I select the starting point $a_{x}$ to be equal to external turning point $a_{\rm tp,\, out}$. Such a choice of starting points turns out to give us higher accuracy in calculations of the total wave function then if to start calculations of both partial solutions from zero or from only one turning point.
So, eq.~(\ref{eq.3.1.3.1}) and furthers in Appendix~\ref{sec.3.1} are defined relatively arbitrary non-zero point $a_{x}$ in a general case. These equations in Appendixes~\ref{sec.3.1} and \ref{sec.3.2} do not absolutely contradict to any desirable boundary condition, and they could be extended on the case $a_{x}=0$. \emph{In latter case the methods in Appendix~B should be working at any finite value of the total wave function at $a=0$.}

In order to obtain the total wave function, we need to connect two partial solutions using one boundary condition, which should be obtained from physical motivations. In the light of logic in Introduction, Sect.~4.1 and 4.2 reinforced by possible interference between the incident and reflected waves in Sect.~5 (which is invisible in the semiclassical approach of the second order) it is more natural not to define the total wave function at zero (or at infinity), but to find outgoing wave at such \underline{finite} value of $a$ in the external region where this wave corresponds to our Universe at present time (as this could be connected with possible astronomical observations). But, practically, it turns out to define more effectively such a wave at point where forces acting on it are as minimal as possible. This is an initial condition imposed on the outgoing wave in the external region\footnote{For example, using such idea for alpha-decay problem in nuclear physics, we obtain the asymptotic region where the outgoing wave is plane (spherical) wave.}.

Now let us analyze a question: which value could the wave function be equal to at zero?
In the paper the following basis is used:
\begin{itemize}
\item
\emph{the wave function should be continuous in the whole spatial region of its definition},

\vspace{-2mm}
\item
\emph{we have outgoing non-zero flux in the asymptotic region defined on the basis of the total wave function},

\vspace{-2mm}
\item
\emph{we consider a case when this flux is constant}.
\end{itemize}
The non-zero outgoing flux defined at arbitrary point requires the wave function to be complex and non-zero. The condition of continuity of this flux in the whole region of definition of the wave function requires this wave function to be complex and non-zero in whole this region. If we included point $a=0$ into the studied region, then we should obtain the non-zero and complex wave function at such a point also. If we use the basis above, then we cannot obtain zero wave function at $a=0$. 
One can pass to nuclear physics where study of such a question and its possible resolutions have longer history then in quantum cosmology. As the simplest demonstration, one can consider elastic scattering of particles on nucleus (where we have zero radial wave function at $r=0$, and we have no any divergences), and alpha decay of nucleus (where we cannot obtain zero wave function at $r=0$).
\emph{A possible divergence of the radial wave function at zero in nuclear decay problem could be explained by existence of source at such a point which creates the outgoing flux in the asymptotic region (and is a reason of such flux).}
Now the picture becomes clearer: any quantum decay could be connected with source at zero. That is why vanishing of the total wave function at $a=0$ after introduction of the wall at this point (like in Ref.~\cite{AcacioDeBarros.2007.PRD}) is not obvious and is only part of all possibilities.


If we wanted to study physics at zero $a=0$, we should come to two cases:
\begin{itemize}
\item
If we include zero point into the region of our consideration, we shall come to quantum mechanics with included sources. In such a case, condition of constant flux is broken. But more general integral formula of non-stationary dependence of the fluxes on probability can include possible sources and put into frameworks of the standard quantum mechanics also (see eq.~(19.5) in Ref.~\cite{Landau.v3.1989}, p.~80). Perhaps, black hole could be another example of quantum mechanics with sources and sinks.

\item
We can consider only quantum mechanics with constant fluxes and without sources. Then we should reduce zero point $a=0$ from the region of our consideration. But the formalism proposed in this paper remains to be working and is able to calculate the penetrability and reflection coefficients without reduce of any accuracy also.
\end{itemize}
This could be \underline{stationary} picture of interdependence between the wave function at zero and the outgoing flux in the asymptotic region. If to pass to non-stationary consideration of this question, then from the problem of possible singularity of the wave function at zero and sources we come to initial condition which should define further evolution of the Universe. In such a case, after defining the initial state (through set of parameters) it is possible to connect zero value of wave packet at $a=0$ (i.~e. without singularity at such a point) with non-zero outgoing flux in the asymptotic region. In such direction different proposals have been made in frameworks of semiclassical models in order to describe inflation, to introduce time or to analyze dynamics of studied quantum system (for example, see \cite{Finelli.1998.PRD,Tronconi.2003.PRD}).

In Ref.~\cite{AcacioDeBarros.2007.PRD} an infinity potential wall at $a=0$ was introduced in order to consider the packets on the positive semiaxis of the scale factor $a$. A question can be appeared: how such an condition could be connected with possibility of the stationary wave functions to be non-zero at $a=0$, according to logics above.
In Appendix~\ref{sec.app.4} penetration of the packet is studied in fully quantum (non-semiclassical) consideration concerning the simplest barrier (which is enough to clarify this question). Here, we present exact analytical solutions for amplitudes. We find coefficients $T_{MIR}$ and $R_{MIR}$ describing penetration of the packet through the barrier from the internal region outside and its reflection, which are separated into the coefficients of penetrability $T_{\rm bar}$ and reflection $R_{\rm bar}$ of the barrier (in standard definition) and new coefficient $\bigl|A_{\rm inc}\bigr|^{2}$ describing oscillations of all multiple packets
inside internal region:
\begin{equation}
\begin{array}{ll}
  T_{MIR} = \bigl|A_{\rm inc}\bigr|^{2} \cdot T_{\rm bar}, &
  R_{MIR} = \bigl|A_{\rm inc}\bigr|^{2} \cdot R_{\rm bar}.
\end{array}
\label{eq.5.5.1}
\end{equation}
This formulas seem to be fully quantum analog of the semiclassical formula of the decay width of a metastable state introduced by \emph{Gurvitz} and \emph{K\"{a}lbermann} in Ref.~\cite{Gurvitz.1987.PRL}, where $\bigl|A_{\rm inc}\bigr|^{2}$ looks to be fully quantum analog of the normalization factor $F$.
In Appendix~\ref{sec.app.4} it is shown that consideration of the packet which penetrates through the barrier with its oscillations inside internal region provides complex stationary total wave function which is non-zero at $a=0$ also (while separate packets can be zero at such a point). So, we obtain full correspondence between non-stationary consideration of the packet penetrated through the barrier and described on the basis of multiple internal reflections concerning boundaries, and the stationary formalism presented in this paper.
Now difference between the penetrability calculated in Ref.~\cite{AcacioDeBarros.2007.PRD} and the penetrability obtained in this paper has became clear:
in Ref.~\cite{AcacioDeBarros.2007.PRD} the reflections of multiple packets are taken from boundary at $a=0$ into account.
(in Ref.~\cite{AcacioDeBarros.2007.PRD} $T_{MIR}$ is obtained, while I calculate $T_{\rm bar}$).
In order to pass from the rectangular barrier considered in Appendix~C to arbitrary barrier shapes, formalism of
Refs.~\cite{Maydanyuk.2003.PhD-thesis,Maydanyuk.arXiv:0805.4165,Maydanyuk.arXiv:0906.4739} could be used, where we present exact solutions concerning barrier consisting from arbitrary finite number of rectangular steps of arbitrary sizes (which is supposed to approximate effectively the studied realistic barrier).

\subsection{The penetrability and reflection: the fully quantum approach versus semiclassical one
\label{sec.6}}

Now we shall estimate by the method described above the coefficients of penetrability and reflection for the potential barrier (\ref{eq.model.5.9}) with parameters $A=36$, $B=12\,\Lambda$, $\Lambda=0.01$ at different values of the energy of radiation $E_{\rm rad}$. We shall compare the found coefficient of penetrability with its value, which the semiclassical method gives.
In the semiclassical approach we shall consider two following definitions of this coefficient:
\begin{equation}
\begin{array}{cc}
  P_{\rm penetrability}^{\rm WKB, (1)} = \displaystyle\frac{1}{\theta^{2}}, &
  P_{\rm penetrability}^{\rm WKB, (2)} = \displaystyle\frac{4}{\Bigl(2\theta + 1/(2\theta)^{2}\Bigr)^{2}},
\end{array}
\label{eq.6.1}
\end{equation}
where
\begin{equation}
  \theta =
    \exp \displaystyle\int\limits_{a_{\rm tp}^{\rm (int)}}^{a_{\rm tp}^{\rm (ext)}} \bigl|V(a)-E\bigr|\; da.
\label{eq.6.2}
\end{equation}
One can estimate also \emph{duration of a formation of the Universe},
using by definition (15) in Ref.~\cite{AcacioDeBarros.2007.PRD}:
\begin{equation}
  \tau = 2\, a_{\rm tp,\, int}\: \displaystyle\frac{1}{\rm P_{penetrability}}.
\label{eq.6.3}
\end{equation}

Results are presented in Tabl.~\ref{table.1} in Appendix 1. In calculations the coefficients of penetrability, reflection and mixing are defined by eqs.~(\ref{eq.4.2.1}), the fluxes by eqs.~(\ref{eq.4.1.3}).
In the whole region of energies up to $E_{\rm rad}=2500$ the calculations give coincidence of the first as a minimum 8 digits between $P_{\rm penetrability}^{\rm WKB, (1)}$ and $P_{\rm penetrability}^{\rm WKB, (2)}$, therefore in the table the results for $P_{\rm penetrability}^{\rm WKB, (1)}$ are included only. From this table one can see that inside whole region of the energy the fully quantum approach gives value for the coefficient of penetrability enough close to its value obtained by the semiclassical approach, that differs essentially from results in the non-stationary approach~\cite{AcacioDeBarros.2007.PRD}. This difference could be explained by difference in a choice of the boundary condition, which is used in construction of the stationary solution of the wave function.

An advantage of the fully quantum method is possibility to calculate the coefficient of reflection. However, the calculations show that this coefficient inside the whole region of the energy used in Tabl.~1 equals to 1 practically. But this coefficient differs visibly from 1 only at the energy of radiation enough close to the height of the barrier (see Tabl.~2 in Appendix), that is explained by essential decreasing of a role of the barrier in penetration of the wave through it. In order to estimate an accuracy of the found coefficients, I obtain property (\ref{eq.4.2.3}) up to the first 11 digits (see Tabl.~2 in Appendix) inside whole region of the energy of radiation. The coefficient of mixing is less than $10^{-19}$ (that is connected with computer error). So, \emph{there is no practical interference between the incident and reflected waves defined by such a way close to the internal turning point that points out their very accurate determination practically.}
Now it becomes clear that the approach proposed in Ref.~\cite{AcacioDeBarros.2007.PRD} and the semiclassical methods do not give such an accuracy in the determination of the coefficients
of penetrability and reflection.

\section{The penetrability in the FRW-model with the Chaplygin gas
\label{sec.7}}

In order to connect the stage of universe with dust matter and its another accelerating stage, in Ref.~\cite{Kamenshchik.2001.PLB} a new scenario with the \emph{Chaplygin gas} was applied to cosmology. A quantum FRW-model with the Chaplygin gas has been constructed on the basis of equation of state instead of $p\,(a)=\rho_{\rm rad}(a)/3$ (where $p\,(a)$ is pressure)
by the following (see also Refs.~\cite{Bilic.2002.PLB,Bento.2002.PRD}):
\begin{equation}
  p_{\rm Ch} = -\displaystyle\frac{A}{\rho_{\rm Ch}^{\alpha}},
\label{eq.7.1.1}
\end{equation}
where $A$ is positive constant and $0< \alpha \le 1$. In particular, for the standard Chaplygin gas we have $\alpha=1$.
Solution of equation of state~(\ref{eq.7.1.1}) gives the following dependence of density on the scale factor:
\begin{equation}
  \rho_{\rm Ch}(a) = \biggl( A + \displaystyle\frac{B}{a^{3\,(1+\alpha)}} \biggr)^{1/(1+\alpha)},
\label{eq.7.1.2}
\end{equation}
where $B$ is a new constant of integration. This model through one phase $\alpha$ connects the stage of Universe where dust dominates and DeSitter stage. Note that for the first time the Chaplygin gas was introduced in aerodynamics~\cite{Chaplygin.1904}.

Let us combine expression for density which includes previous forms of matter and the Chaplygin gas in addition. At limit $\alpha \to 0$ eq.~(\ref{eq.7.1.2}) is transformed into the $\rho_{\rm dust}$ component plus the $\rho_{\Lambda}$ component.
From such limit we find
\begin{equation}
\begin{array}{cc}
  A = \rho_{\Lambda}, &
  B = \rho_{\rm dust}
\end{array}
\label{eq.7.1.3}
\end{equation}
and obtain the following generalized density:
\begin{equation}
  \rho\,(a) =
    \biggl( \rho_{\Lambda} + \displaystyle\frac{\rho_{\rm dust}}{a^{3\,(1+\alpha)}} \biggr)^{1/(1+\alpha)} +
    \displaystyle\frac{\rho_{\rm rad}}{a^{4}(t)}.
\label{eq.7.1.4}
\end{equation}
Now we have:
\begin{equation}
  \dot{a}^{2} + k -
  \displaystyle\frac{8\pi\,G}{3}\,
    \Biggl\{
      a^{2}\:\biggl( \rho_{\Lambda} + \displaystyle\frac{\rho_{\rm dust}}{a^{3\,(1+\alpha)}} \biggr)^{1/(1+\alpha)} +
      \displaystyle\frac{\rho_{\rm rad}}{a^{2}(t)}
    \Biggr\}= 0.
\label{eq.7.1.5}
\end{equation}
After quantization we obtain the Wheeler-De Witt equation
\begin{equation}
\begin{array}{cc}
  \biggl\{ -\:\displaystyle\frac{\partial^{2}}{\partial a^{2}} + V_{\rm Ch}\,(a) \biggr\}\; \varphi(a) =
  E_{\rm rad}\; \varphi(a), &
  E_{\rm rad} = \displaystyle\frac{3\,\rho_{\rm rad}}{2\pi\,G},
\end{array}
\label{eq.7.2.1}
\end{equation}
where
\begin{equation}
\begin{array}{ccl}
  V_{\rm Ch}\,(a) & = &
    \biggl( \displaystyle\frac{3}{4\pi\,G} \biggr)^{2}\: k\,a^{2} -
    \displaystyle\frac{3}{2\pi\,G}\:
     a^{4}\, \biggl( \rho_{\Lambda} + \displaystyle\frac{\rho_{\rm dust}}{a^{3\,(1+\alpha)}} \biggr)^{1/(1+\alpha)}.
\end{array}
\label{eq.7.2.2}
\end{equation}
For the Universe of closed type (at $k=1$) at $8\pi\,G \equiv M_{\rm p} = 1$ we have
(see eqs.~(6)--(7) in Ref.~\cite{Bouhmadi-Lopez.2005.PRD}):
\begin{equation}
\begin{array}{cc}
  V_{\rm Ch}\,(a) =
    36\,a^{2} -
    12\,a^{4}\,\Bigl(\Lambda + \displaystyle\frac{\rho_{\rm dust}}{a^{3\,(1+\alpha)}} \Bigr)^{1/(1+\alpha)}, &
    E_{\rm rad} = 12\, \rho_{\rm rad}.
\end{array}
\label{eq.7.2.3}
\end{equation}
\begin{figure}[h]
\centering{\includegraphics[width=7cm]{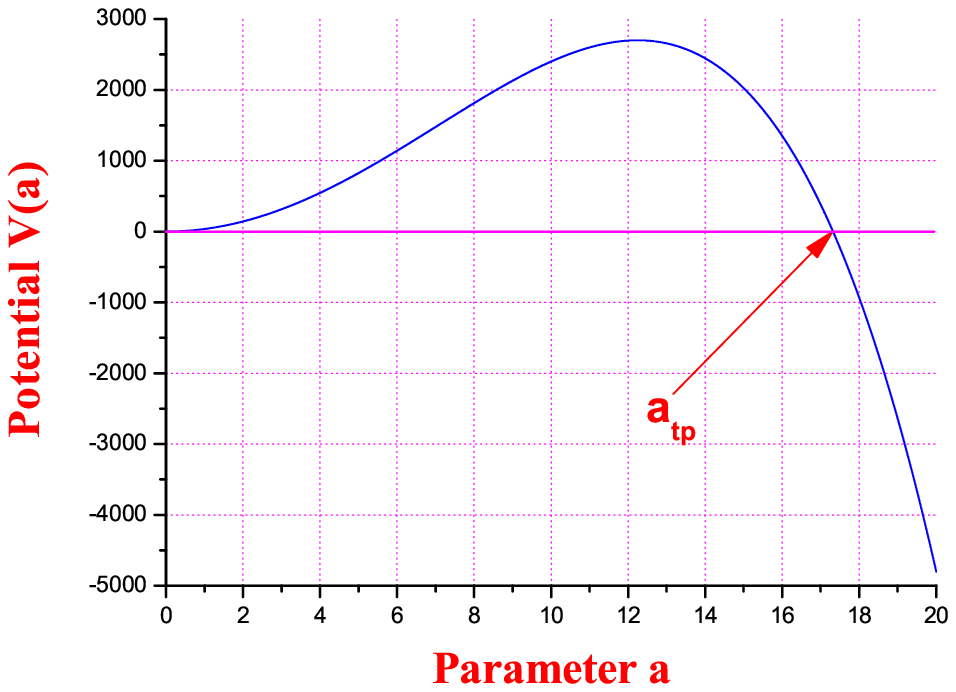}
\includegraphics[width=7cm]{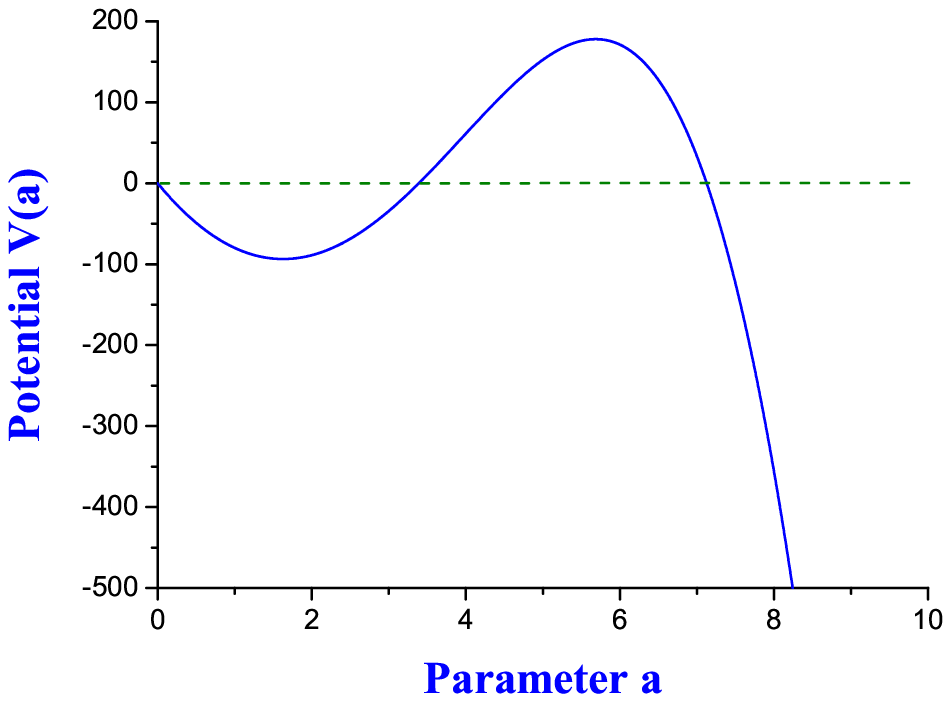}}
\caption{Cosmological potentials with and without Chaplygin gas:
Left panel is for potential $V(a) = 36\,a^{2} - 12\,\Lambda\,a^{4}$ with parameter $\Lambda=0.01$ (turning point $a_{tp} = 17.320508$ at zero energy $E_{\rm rad}=0$),
Right panel is for potential (\ref{eq.7.2.3}) with parameters $\Lambda=0.01$, $\rho_{\rm dust}=30$, $\alpha=0.5$
(minimum of the hole is -93.579 and its coordinate is 1.6262,
maximum of the barrier is 177.99 and its coordinate is 5.6866).
\label{fig.8}}
\end{figure}

Let us expand the potential (\ref{eq.7.2.3}) close to arbitrary selected point $\bar{a}$ by powers $q=a-\bar{a}$
and restrict ourselves by linear item only:
\begin{equation}
  V_{\rm Ch}\,(q) = V_{0} + V_{1}q.
\label{eq.7.3.1}
\end{equation}
For coefficients $V_{0}$ and $V_{1}$ I find:
\begin{equation}
\begin{array}{ccl}
  V_{0} & = & V_{\rm Ch}\,(a=\bar{a}), \\
  V_{1} & = &
    \displaystyle\frac{d V_{\rm Ch}\,(a)}{da} \bigg|_{a=\bar{a}} =
    72\,a +
    12\,a^{3}\,
    \Bigl\{ -4\,\Lambda - \displaystyle\frac{\rho_{\rm dust}}{a^{3\,(1+\alpha)}} \Bigr\} \cdot
    \Bigl( \Lambda + \displaystyle\frac{\rho_{\rm dust}}{a^{3\,(1+\alpha)}} \Bigr)^{-\alpha/(1+\alpha)}
\end{array}
\label{eq.7.3.3}
\end{equation}
and eq.~(\ref{eq.7.2.1}) obtains a form:
\begin{equation}
  -\displaystyle\frac{d^{2}}{dq^{2}}\, \varphi(q) + (V_{0} - E_{\rm rad} + V_{1}\, q)\: \varphi(q) = 0.
\label{eq.7.3.4}
\end{equation}
After change of variable
\begin{equation}
\begin{array}{cc}
  \zeta = |V_{1}|^{1/3}\, q, &
  \displaystyle\frac{d^{2}}{dq^{2}} =
  \Bigl(\displaystyle\frac{d\zeta}{dq} \Bigr)^{2}\, \displaystyle\frac{d^{2}}{d\zeta^{2}} =
  |V_{1}|^{2/3}\; \displaystyle\frac{d^{2}}{d\zeta^{2}}
\end{array}
\label{eq.7.3.6}
\end{equation}
eq.~(\ref{eq.7.3.4}) is transformed into the following:
\begin{equation}
  \displaystyle\frac{d^{2}}{d\zeta^{2}}\, \varphi(\zeta) +
  \biggl\{\displaystyle\frac{E_{\rm rad} - V_{0}} {|V_{1}|^{2/3}} -
          \displaystyle\frac{V_{1}} {|V_{1}|}\: \zeta\biggr\}\: \varphi(\zeta) = 0.
\label{eq.7.3.7}
\end{equation}
After new change
\begin{equation}
  \xi = \displaystyle\frac{E_{\rm rad} - V_{0}} {|V_{1}|^{2/3}} - \displaystyle\frac{V_{1}} {|V_{1}|}\: \zeta
\label{eq.7.3.8}
\end{equation}
we have
\begin{equation}
  \displaystyle\frac{d^{2}}{d\xi^{2}}\, \varphi(\xi) + \xi\, \varphi(\xi) = 0.
\label{eq.7.3.9}
\end{equation}
From eqs.~(\ref{eq.7.3.6}) and (\ref{eq.7.3.8}) we write:
\begin{equation}
  \xi = \displaystyle\frac{E_{\rm rad} - V_{0}} {|V_{1}|^{2/3}} - \displaystyle\frac{V_{1}} {|V_{1}|^{2/3}}\: q.
\label{eq.7.3.10}
\end{equation}

Using such corrections after inclusion of the density component of the Chaplygin gas, I have calculated the wave function and on its basis the coefficients of penetrability, reflection and mixing by the formalism presented above.
Now according to the logic of Sec.~\ref{sec.5.1}, I have defined the incident and reflected waves relatively new boundary which is located in the minimum of the hole in the internal region. Results are presented in Tabl.~3. One can see that penetrability is changed up to 100 times in dependence on the location of the boundary in such a coordinate or in the internal turning point (for the same fixed barrier shape and energy $E_{\rm rad}$)! This confirms that coordinate where incident and reflected waves are defined has essential influence on estimation of the coefficients of penetrability and reflection. This result shows real physical sense of the method proposed in the given paper. In the next Tabl.~4 I demonstrate fulfillment of the property (\ref{eq.4.2.3}) inside the whole energy region, which is calculated on the basis of the coefficients of penetrability, reflection and mixing obtained before. One can see that accuracy is the first 10--12 digits. Of course, any semiclassical calculations are not able to give such accuracy for penetrability for the studied cosmological barriers.


\section{Conclusions and perspectives
\label{sec.conclusions}}

In the paper the closed Friedmann--Robertson--Walker model with quantization in the presence of a positive cosmological constant and radiation is studied. I have solved it numerically and have determined the tunneling probability for the birth of an asymptotically deSitter, inflationary Universe as a function of the radiation energy.
Note the following.
\begin{enumerate}
\item
A formalism for calculation of two linear independent partial solutions for the wave function of the Universe for the scale factor inside the region $0 \le a \le 100$ and the energy of radiation from zero up to the barrier height has been constructed.

\item
A fully quantum definition of the wave which propagates inside strong field and interact minimally with them, has been formulated for the first time, and approach for its stable determination has been constructed.

\item
A new stationary approach for determination of the incident, reflected and transmitted waves relatively a barrier has been constructed, the tunneling boundary condition has been corrected.

\item
A quantum stationary method of determination of coefficients of penetrability and reflection relatively the barrier with analysis of uniqueness of solution has been developed, where for the first time non-zero interference between the incident and reflected waves has been taken into account and for its estimation the coefficient of mixing has been introduced.

\item
A criterion of estimation of accuracy of the determination of these coefficients has been proposed on the basis of check of eq.~(\ref{eq.4.2.3}).
\end{enumerate}
In such a quantum approach the penetrability of the barrier for the studied quantum cosmological model with $A=36$, $B=12\,\Lambda$ parameters at $\Lambda=0.01$ has been estimated with a comparison with results of other known methods.
Note the following.
\begin{itemize}
\item
According to the calculations, inside whole region of energy of radiation the tunneling probability for the birth of an asymptotically deSitter, inflationary Universe is very close to its value, obtained in the semiclassical approach by eqs.~(\ref{eq.6.1}) and (\ref{eq.6.2}), but essentially differs on the results obtained before by the quantum non-stationary approach in Ref.~\cite{AcacioDeBarros.2007.PRD} (see Tabl.~1 and 2 in Appendix).

\item
The coefficient of reflection from the barrier in the internal region has been determined at first time. According to calculations, it is differed essentially on 1 at the energy of radiation close enough to the barrier height (see Tabl.~2 in Appendix).

\item
The modulus of the coefficient of mixing is less $10^{-19}$ for all energies, that points out that \emph{there is no \underline{interference} between the found incident and reflected waves close to the internal turning point.}

\item
On the basis of the calculated coefficients I reconstruct a property (\ref{eq.4.2.3}) with accuracy of the first 11--18 digits (see Tabl.~2 in Appendix) inside the whole studied region of the energy of radiation.
\end{itemize}
One can assume that the method proposed can be easily generalized on the other cosmological models with the barriers with arbitrary complicated shapes. However, one can suppose that a visible change can be found in difference between the penetrabilities in the fully quantum and semiclassical approaches after inclusion of new components of density into the model (for example, after use of the component of Chaplygin gas in
Refs.~\cite{Chaplygin.1904,Kamenshchik.2001.PLB,Bilic.2002.PLB,Bento.2002.PRD}).
Of course, it can be interesting to compare the estimations of the coefficients of penetrability and reflection with results which could be obtained on the basis of the \emph{Improved WKB approach} on the basis of Refs.~\cite{Casadio.2005.PRD.D71,Casadio.2005.PRD.D72,Luzzi.2006.PhD}.


\appendix
\section{Cosmological model in the Friedmann--Robertson--Walker metric
\label{sec.model}}

\subsection{Dynamics of Universe in the Friedmann--Robertson--Walker metric
\label{sec.app.model}}

Let us consider a simple model of the homogeneous and isotropic Universe in \emph{Friedmann--Robertson--Walker (FRW) metric} (see Ref.~\cite{Weinberg.1975}, p.~438; also see
Refs.~\cite{Rubakov.RTN2005,Linde.2005,Trodden.TASI-2003,Brandenberger.1999}):
\begin{equation}
\begin{array}{cccccc}
  ds^{2} = - dt^{2} + a^{2}(t) \cdot \biggl(\displaystyle\frac{dr^{2}}{h(r)} + r^{2} (d\theta^{2} + \sin^{2}{\theta} \, d\phi^{2}) \biggr), &
  h(r) = 1-kr^{2},
\end{array}
\label{eq.app.model.1.1}
\end{equation}
where $t$ and $r$, $\theta$, $\phi$ are time and space spherical coordinates,
the signature of the metric is $(-,+,+,+)$ as in Ref.~\cite{Trodden.TASI-2003} (see~p.~4),
$a(t)$ is an unknown function of time and $k$ is a constant, the value of which equals $+1$, $0$ or $-1$, with appropriate choice of units for $r$. Further, we shall use the following system of units: $\hbar = c = 1$.
For $k = -1$, 0 the space is infinite (Universe of open type), and for $k=+1$ the space is finite (the Universe of closed type). For $k=1$ one can describe the space as a sphere with radius $a(t)$ embedded in a 4-dimensional Euclidian space. The function $a(t)$ is referred to as the \emph{``radius of the Universe''} and is called the \emph{cosmic scale factor}. This function contains information of the dynamics of the expansion of the Universe, and therefore its determination is an actual task.

One can find the function $a(t)$ using the Einstein equations with taking into account of the cosmological constant $\Lambda$ in this metric (we use the signs according to the chosen signature, as in Ref.~\cite{Trodden.TASI-2003} p.~8; the Greek symbols $\mu$ and $\nu$ denote any of the four coordinates $t$, $r$, $\theta$ and $\phi$):
\begin{equation}
  R_{\mu\nu} - \displaystyle\frac{1}{2} \, g_{\mu\nu} \, R = 8\pi \: G \, T_{\mu\nu} + \Lambda,
\label{eq.app.model.1.2}
\end{equation}
where $R_{\mu\nu}$ is the Ricci tensor, $R$ is the scalar curvature, $T_{\mu\nu}$ is the energy-momentum tensor, and $G$ is Newton's constant.
From (\ref{eq.app.model.1.1}) we find the Ricci tensor $R_{\mu\nu}$ and the scalar curvature $R$:
\begin{equation}
\begin{array}{ll}
  \vspace{2mm}
  R_{tt} = -3 \displaystyle\frac{\ddot{a}}{a}, &
  \hspace{10mm}
  R_{rr} = \displaystyle\frac{a\ddot{a}}{h} +2\displaystyle\frac{\dot{a}^{2}}{h} -\displaystyle\frac{h^{\prime}}{hr} = \displaystyle\frac{2\dot{a}^{2} + a\ddot{a} + 2k}{1-kr^{2}}, \\
  R_{\phi\phi} = R_{\theta\theta} \, \sin^{2}{\theta}, &
  \hspace{10mm}
  R_{\theta\theta} = a\ddot{a}\,r^{2} + 2\dot{a}^{2}\,r^{2} - h - \displaystyle\frac{h^{\prime}r}{2} + 1 = 2\dot{a}^{2}\,r^{2} + a\ddot{a}^{2}\,r^{2} + 2kr^{2}
\end{array}
\label{eq.app.model.1.12}
\end{equation}
\begin{equation}
  R = g^{tt} R_{tt} + g^{rr} R_{rr} + g^{\theta\theta} R_{\theta\theta} + g^{\phi\phi} R_{\phi\phi} =
  \displaystyle\frac{6\dot{a}^{2} + 6a\ddot{a} + 6k}{a^{2}}.
\label{eq.app.model.1.14}
\end{equation}

The \emph{energy-momentum tensor} has a form (see~\cite{Trodden.TASI-2003}, p.~8):
$T_{\mu\nu} = (\rho + p) \: U_{\mu} U_{\nu} + p \: g_{\mu\nu}$,
where $\rho$ and $p$ are energy density and pressure.
Here, one needs to use the normalized vector of 4-velocity $U^{t} = 1$, $U^{r} = U^{\theta} = U^{\phi} = 0$.
%
%
Substituting the found components (\ref{eq.app.model.1.12}) of the Ricci tensor $R_{\mu\nu}$, the scalar curvature (\ref{eq.app.model.1.14}), the components of the energy-momentum tensor $T_{\mu\nu}$ and including the component $\rho_{\rm rad}(a)$, describing the radiation in the initial stage (equation of state for radiation: $p(a)=\rho_{\rm rad}(a)/3$), into the Einstein's equation (\ref{eq.app.model.1.2}) at $\mu=\nu=0$,
we obtain the \emph{Friedmann equation} with taking into account the cosmological constant (see p.~8 in Ref.~\cite{Trodden.TASI-2003}; p.~3 in Ref.~\cite{Brandenberger.1999}; p.~2 in Ref.~\cite{Vilenkin.1995}):
\begin{equation}
\begin{array}{ll}
  \dot{a}^{2} + k -
  \displaystyle\frac{8\pi\,G}{3}\,
  \Bigl\{\displaystyle\frac{\rho_{\rm rad}}{a^{2}(t)} + \rho_{\Lambda}\,a^{2}(t) \Bigr\} = 0, &
  \hspace{5mm}
  \rho_{\Lambda} = \displaystyle\frac{\Lambda}{8\pi\,G},
\end{array}
\label{eq.app.model.2.2.5}
\end{equation}
where $\dot{a}$ is derivative $a$ at time coordinate.
From here, we write a general expression for the energy density:
\begin{equation}
  \rho\,(a) = \rho_{\Lambda} + \displaystyle\frac{\rho_{\rm rad}}{a^{4}(t)}.
\label{eq.app.model.2.2.8}
\end{equation}

\subsection{Action, lagrangian and quantization
\label{sec.model.3}}

We define the action as in Ref.~\cite{Vilenkin.1995} (see~(1), p.~2):
\begin{equation}
  S = \displaystyle\int \sqrt{-g}\: \biggl( \displaystyle\frac{R}{16\pi\,G} - \rho \biggr)\; dx^{4}.
\label{eq.app.model.3.1}
\end{equation}
Substituting the scalar curvature (\ref{eq.app.model.1.14}), then integrating item at $\ddot{a}$ by parts with respect to variable $t$, we obtain the \emph{lagrangian} (see Ref.~\cite{Vilenkin.1995}, (11), p.~4):
\begin{equation}
  \mathcal{L}\,(a,\dot{a}) =
  \displaystyle\frac{3\,a}{8\pi\,G}\:
  \biggl(-\dot{a}^{2} + k - \displaystyle\frac{8\pi\,G}{3}\; a^{2}\,\rho(a) \biggr).
\label{eq.app.model.3.8}
\end{equation}
Considering the variables $a$ and $\dot{a}$ as generalized coordinate and velocity respectively, we find a generalized momentum conjugate to $a$:
\begin{equation}
  p_{a} = \displaystyle\frac{\partial\, \mathcal{L}\,(a,\dot{a})}{\partial \dot{a}} =
  - \,\displaystyle\frac{3}{4\pi\,G}\; a\,\dot{a}
\label{eq.app.model.4.1}
\end{equation}
and then hamiltonian:
\begin{equation}
\begin{array}{ccl}
  \vspace{2mm}
  h\,(a,p_{a}) & = & p\,\dot{a} - \mathcal{L}\,(a,\dot{a}) =
  -\:\displaystyle\frac{1}{a}\;
  \biggl\{
    \displaystyle\frac{2\pi\,G}{3}\: p_{a}^{2} +
    a^{2}\,\displaystyle\frac{3\,k}{8\pi\,G} -
    a^{4}\,\rho(a) \biggr\}.
\end{array}
\label{eq.app.model.4.2}
\end{equation}
The passage to the quantum description of the evolution of the Universe is obtained by the standard procedure of canonical quantization in the Dirac formalism for systems with constraints. In result, we obtain the \emph{Wheeler--De Witt (WDW) equation} (see Ref.~\cite{Vilenkin.1995}, (16)--(17), in p.~4, \cite{Wheeler.1968,DeWitt.1967,Rubakov.2002.PRD}), which after multiplication on factor and passage of the item at the component with radiation $\rho_{\rm rad}$ into right part transforms into the following form:
\begin{equation}
\begin{array}{ccl}
  \biggl\{ -\:\displaystyle\frac{\partial^{2}}{\partial a^{2}} + V\,(a) \biggr\}\; \varphi(a) =
  E_{\rm rad}\; \varphi(a), &
  V\, (a) =
    \biggl( \displaystyle\frac{3}{4\pi\,G} \biggr)^{2}\: k\,a^{2} -
    \displaystyle\frac{3\,\rho_{\Lambda}}{2\pi\,G}\; a^{4}, &
  E_{\rm rad} = \displaystyle\frac{3\,\rho_{\rm rad}}{2\pi\,G}.
\end{array}
\label{eq.app.model.5.3}
\end{equation}
This equation looks similar to the one-dimensional stationary Schr\"{o}dinger equation on a semiaxis (of the variable $a$) at energy $E_{\rm rad}$ with potential $V\,(a)$. For further analysis one can convenient to use the system of units where $8\pi\,G \equiv M_{\rm p} = 1$.
Let us rewrite $V\,(a)$ in a generalized form:
\begin{equation}
  V(a) = A\,a^{2} - B\,a^{4}.
\label{eq.app.model.5.9}
\end{equation}
In particular, for the Universe of the closed type ($k=1$) we obtain $A = 36$, $B = 12\,\Lambda$ (this potential coincides with \cite{AcacioDeBarros.2007.PRD}).


\subsection{Potential close to the turning points: non-zero energy case
\label{model.6}}

At first, let us find the \emph{turning points} $a_{\rm tp,\,in}$ and $a_{\rm tp,\,out}$ concerning the potential (\ref{eq.app.model.5.9}) at energy $E_{\rm rad}$:
\begin{equation}
\begin{array}{cc}
\vspace{3mm}
  a_{\rm tp,\, in} =
    \sqrt{\displaystyle\frac{A}{2B}} \cdot
    \sqrt{1 - \sqrt{1 - \displaystyle\frac{4BE_{\rm rad}}{A^{2}}}},&
  a_{\rm tp,\, out} =
    \sqrt{\displaystyle\frac{A}{2B}} \cdot
    \sqrt{1 + \sqrt{1 - \displaystyle\frac{4BE_{\rm rad}}{A^{2}}}}.
\end{array}
\label{eq.app.model.6.2}
\end{equation}
Let us expand the potential $V(a)$ (\ref{eq.app.model.5.9}) in powers $q_{\rm out}=a-a_{\rm tp}$ (where as $a_{\rm tp}$ the point $a_{\rm tp,\, in}$ or $a_{\rm tp,\, out}$ is used, close to which we find expansion),
where (for small $q$) we restrict ourselves by the linear item only:
\begin{equation}
  V(q) = V_{0} + V_{1},
\label{eq.app.model.6.3}
\end{equation}
where the coefficients $V_{0}$ and $V_{1}$ are:
\begin{equation}
\begin{array}{lcl}
\vspace{1mm}
  V_{0} & = &
    V(a=a_{\rm tp,\, in}) =
    V(a=a_{\rm tp,\, out}) =
    A\, a_{\rm tp}^{2} - B\, a_{\rm tp}^{4} = E_{\rm rad}, \\
\vspace{1mm}
  V_{1}^{\rm (out)} & = &
    -\: 2\, A \cdot
    \sqrt{\displaystyle\frac{A}{2B}\:
    \biggl(1 - \displaystyle\frac{4BE_{\rm rad}}{A^{2}}\biggr)\,
    \biggl(1 + \sqrt{1 - \displaystyle\frac{4BE_{\rm rad}}{A^{2}}}\biggr)}, \\
  V_{1}^{\rm (int)} & = &
    2\,A \cdot
    \sqrt{\displaystyle\frac{A}{2B}\:
    \biggl(1 - \displaystyle\frac{4BE_{\rm rad}}{A^{2}}\biggr)\,
    \biggl(1 - \sqrt{1 - \displaystyle\frac{4BE_{\rm rad}}{A^{2}}}\biggr)}.
\end{array}
\label{eq.app.model.6.5}
\end{equation}
Now eq.~(\ref{eq.app.model.5.3}) transforms into a new form at variable $q$ with potential $V(q)$:
\begin{equation}
  -\displaystyle\frac{d^{2}}{dq^{2}}\, \varphi(q) +
  V_{1}\, q\: \varphi(q) = 0.
\label{eq.app.model.6.6}
\end{equation}

\section{Calculations of the wave function of Universe
\label{sec.app.3}}

\subsection{Method of beginning of the solution
\label{sec.3.1}}

We shall be looking for the regular partial solution for the wave function close to arbitrary selected point $a_{x}$.
Let us write the wave function in the form:
\begin{equation}
\begin{array}{ll}
  \varphi(a) = c_{2} \sum\limits_{n=0}^{+\infty} b_{n} \: (a-a_{x})^{n}
  = c_{2} \sum\limits_{n=0}^{+\infty} b_{n} \: \bar{a}^{n}, &
  \bar{a} = a-a_{x}
\end{array}
\label{eq.3.1.3.1}
\end{equation}
and rewrite the potential through the variable $\bar{a}$:
\begin{equation}
  V(a) = C_{0} + C_{1}\,\bar{a} + C_{2}\,\bar{a}^{2} + C_{3}\,\bar{a}^{3} + C_{4}\,\bar{a}^{4},
\label{eq.3.1.3.4}
\end{equation}
where
\begin{equation}
\begin{array}{ccl}
  C_{0} & = & A\, a_{x}^{2} - B \,a_{x}^{4}, \\
  C_{1} & = & 2a_{x}(A-B\,a_{x}^{2}) - 2B\,a_{x}^{3} = 2A\,a_{x} - 4B\,a_{x}^{3}, \\
  C_{2} & = & A - B\,a_{x}^{2} - 4B\, a_{x}^{2} - B\,a_{x}^{2} = A - 6B\,a_{x}^{2}, \\
  C_{3} & = & -2B\,a_{x} - 2B\,a_{x} = -4B\,a_{x}, \\
  C_{4} & = & - B.
\end{array}
\label{eq.3.1.3.5}
\end{equation}
Substituting the wave function (\ref{eq.3.1.3.1}), its second derivative and the potential (\ref{eq.3.1.3.4}) into Schr\"{o}dinger equation,
we obtain recurrent relations for calculation of the unknown $b_{n}$:
\begin{equation}
\begin{array}{ccccrl}
  b_{2} = \displaystyle\frac{(C_{0}-E)\,b_{0}}{2}, &
  b_{3} = \displaystyle\frac{(C_{0}-E)\,b_{1} + C_{1}\,b_{0}}{6}, &
  b_{4} = \displaystyle\frac{(C_{0}-E)\,b_{2} + C_{1}\,b_{1} + C_{2}\,b_{0}}{12}, &
\end{array}
\label{eq.3.1.3.8}
\end{equation}
\begin{equation}
\begin{array}{ccccrl}
  b_{5} = \displaystyle\frac{(C_{0}-E)\,b_{3} + C_{1}\,b_{2} + C_{2}\,b_{1} + C_{3}\,b_{0}}{20},
\end{array}
\label{eq.3.1.3.9}
\end{equation}
\begin{equation}
\begin{array}{ccccrl}
  b_{n+2} =
  \displaystyle\frac{(C_{0}-E)\,b_{n}+ C_{1}\,b_{n-1}+ C_{2}\,b_{n-2}+ C_{3}\,b_{n-3}+ C_{4}\,b_{n-4}}
  {(n+1)\,(n+2)} & \mbox{at } n \ge 4.
\end{array}
\label{eq.3.1.3.10}
\end{equation}
For given values for $b_{0}$ and $b_{1}$ using eqs.~(\ref{eq.3.1.3.8})--(\ref{eq.3.1.3.10}) one can calculate all $b_{n}$ needed. At limit $E_{\rm rad} \to 0$ and at $a_{x} = 0$ all found solutions for $b_{i}$ are transformed into the corresponding solutions (40), early obtained in \cite{Maydanyuk.2008.EPJC} at $E_{\rm rad}=0$.
Using $c_{2}=1$, from eqs.~(\ref{eq.3.1.3.1}) we find:
\begin{equation}
\begin{array}{cc}
  b_{0} = \varphi\,(a_{x}), & b_{1} = \varphi^{\prime}(a_{x}).
\end{array}
\label{eq.3.1.3.11}
\end{equation}
So, on the basis of the coefficients $b_{0}$ and $b_{1}$ one can obtain values for the wave function and its derivative at point $a_{x}$. Implying two different boundary conditions through $b_{0}$ and $b_{1}$ (in such a way that they locate the first node for the wave function at different places), we obtain two \emph{linearly independent partial solutions $\varphi_{1}(a)$ and $\varphi_{2}(a)$ for the wave function}. Using the internal turning point $a_{\rm tp,\,in}$ as the starting point, by such a way we calculate the first partial solution which increases in the barrier region (we select: $b_{0} = 0.1$, $b_{1} = 1$), and using the external turning point $a_{\rm tp,\,out}$ as the starting point, by such a way we calculate the second partial solution which decreases in the barrier region (we select: $b_{0} = 1$, $b_{1} = -0.1$). Such a choice provides effectively a linear independence between two partial solutions.


\subsection{Method of continuation of the solution
\label{sec.3.2}}

Let us rewrite eq.~(\ref{eq.app.model.5.3})
in such a form\footnote{in obtaining this algorithm a logics from \cite{Zaichenko_Kashuba.2001} was used}:
\begin{equation}
  \varphi^{\prime\prime}\,(a) = f\,(a)\: \varphi\,(a).
\label{eq.3.2.1}
\end{equation}
Let $\bigl\{a_{n} \bigr\}$ be a set of equidistant points $a_{n} = a_{0} + nh$. Denoting values of the wave function $\varphi\,(a)$ at points $a_{n}$ as $\varphi_{n}$, we have constructed own algorithm of the ninth order for
determining $\varphi_{n+1}$ and $\varphi_{n}^{\prime}$ on the previously known $\varphi_{n}$ and $\varphi_{n-1}$:
\begin{equation}
\begin{array}{ccl}
  \vspace{1mm}
  \varphi_{n+1} & = &
    \varphi_{n-1}\:\displaystyle\frac{g_{11} + g_{01}}{g_{01} - g_{11}} +
    \varphi_{n}\:\displaystyle\frac{g_{01}\,g_{10} - g_{00}\,g_{11}}{g_{01} - g_{11}} + O\,(h^{9}), \\

  \varphi_{n}^{\prime} & = &
    \varphi_{n-1}\:\displaystyle\frac{2}{g_{01} - g_{11}} +
    \varphi_{n}\:\displaystyle\frac{g_{10} - g_{00}}{g_{01} - g_{11}} + O\,(h^{9}),
\end{array}
\label{eq.3.2.8}
\end{equation}
where
\begin{equation}
\begin{array}{ccl}
\vspace{1mm}
  g_{00} & = &
    2  + h^{2}\,f_{n} +
    \displaystyle\frac{2}{4!}\: h^{4}\, \bigl(f_{n}^{\prime\prime} + f_{n}^{2}\bigr) +
    \displaystyle\frac{2}{6!}\: h^{6}\,
      \Bigl(f_{n}^{(4)} + 4\,\bigl(f_{n}^{\prime}\bigr)^{2} + 7\,f_{n}\,f_{n}^{\prime\prime} + f_{n}^{3}\Bigr) + \\
\vspace{3mm}
    & + &
    \displaystyle\frac{2}{8!}\: h^{8}\,
      \Bigl(
        f_{n}^{(6)} + 16\,f_{n}\,f_{n}^{(4)} + 26\,f_{n}^{\prime}\,f_{n}^{(3)} +
        15\,\bigl(f_{n}^{\prime\prime}\bigr)^{2} +
        22\,f_{n}^{2}\,f_{n}^{\prime\prime} +
        28\,f_{n}\,\bigl(f_{n}^{\prime}\bigr)^{2} + f_{n}^{4}
      \Bigr), \\

\vspace{2mm}
  g_{01} & = &
    \displaystyle\frac{2}{4!}\: h^{4}\, 2\,f_{n}^{\prime} +
    \displaystyle\frac{2}{6!}\: h^{6}\,
      \Bigl(4\,f_{n}^{(3)} + 6\,f_{n}\,f_{n}^{\prime} \Bigr) +
    \displaystyle\frac{2}{8!}\: h^{8}\,
      \Bigl(
        6\,f_{n}^{(5)} + 24\,f_{n}\,f_{n}^{(3)} + 48\,f_{n}^{\prime}f_{n}^{\prime\prime} +
        12\,f_{n}^{2}\,f_{n}^{\prime}
      \Bigr), \\

\vspace{2mm}
  g_{10} & = &
    \displaystyle\frac{2}{3!}\: h^{3}\,f_{n}^{\prime} +
    \displaystyle\frac{2}{5!}\: h^{5}\,\bigl(f_{n}^{(3)} + 4\,f_{n}\,f_{n}^{\prime}\bigr) +
    \displaystyle\frac{2}{7!}\: h^{7}\,
      \Bigl( f_{n}^{(5)} + 11\,f_{n}\,f_{n}^{(3)} +
             15\,f_{n}^{\prime}f_{n}^{\prime\prime} + 9\,f_{n}^{2}\,f_{n}^{\prime} \Bigr), \\

  g_{11} & = &
    2\,h +
    \displaystyle\frac{2}{3!}\:h^{3}\,f_{n} +
    \displaystyle\frac{2}{5!}\:h^{5}\, \bigl(3\,f_{n}^{\prime\prime} + f_{n}^{2}\bigr) +
    \displaystyle\frac{2}{7!}\: h^{7}\,
      \Bigl(5\,f_{n}^{(4)} + 13\,f_{n}\,f_{n}^{\prime\prime} + 10\,\bigl(f_{n}^{\prime}\bigr)^{2} + f_{n}^{3} \Bigr).
\end{array}
\label{eq.3.2.7}
\end{equation}
A local error of these formulas at point $a_{n}$ equals to:
\begin{equation}
  \delta_{n} =
    \displaystyle\frac{1}{10!}\: h^{10}\, f_{n}^{\prime}\, \varphi_{n}^{(7)}.
\label{eq.3.2.9}
\end{equation}

\section{Tunneling of the packet through radial rectangular barrier
\label{sec.app.4}}

Let us consider a problem of quantum tunneling of the packet through the barrier used in cosmological model.
We shall study such a process consequently by steps of its propagation relatively to each boundary of the barrier, using developed formalism of multiple internal reflections presented in
Refs.~\cite{Maydanyuk.2000.UPJ,Maydanyuk.2002.JPS,Maydanyuk.2002.PAST,Maydanyuk.2003.PhD-thesis,%
Maydanyuk.2006.FPL,Maydanyuk.arXiv:0805.4165,Maydanyuk.arXiv:0906.4739}
(see also Refs.~\cite{Fermor.1966.AJPIA,McVoy.1967.RMPHA,Anderson.1989.AJPIA,Esposito.2003.PRE}).
In order to form idea of multiple internal reflections of the packets in description of quantum tunneling on the positive semiaxis of the scale factor $a$, we shall use the simplest potential $V(a)$:
$V(a)=-V_{0}$ for $0 < a < R_{1}$ (internal region I),
$V(a)=V_{1}$ for $R_{1} < a < R_{2}$ (region II of the barrier) and
$V(a)=0$ for $a > R_{2}$ (external region III). 
%
%
For simplicity, we start from consideration of the case when total energy of system $E$ is higher then the barrier height $V_{1}$: $E>V_{1}$.

In the first step we consider the packet in the region I propagating to the right,
which is incident on the first boundary of the barrier at $R_{1}$:
\begin{equation}
\begin{array}{lcll}
  \psi_{\rm inc}^{(1)}(a, t) & = &
    \int\limits_{E_{\rm min}}^{+\infty} g(E - \bar{E})\: e^{ik_{1}a -iEt/\hbar}\; dE
    & \mbox{at } 0<a<R_{1},
\end{array}
\label{eq.app.4.1}
\end{equation}
where $k_{1} = \sqrt{E+V_{0}}$, $E$ is the energy.
The weight amplitude $g(E - \bar{E})$ can be used in standard form of gaussian and satisfies to normalization $\int |g(E - \bar{E})|^{2}\: dE = 1$, value $\bar{E}$ is an average energy.
This packet transforms into two new packets: the first packet transmitted through this boundary and propagating further in the region II,
and the second one reflected from the boundary and propagating back in the region I:
\begin{equation}
\begin{array}{lcll}
  \psi^{(1)}_{\rm tr}(a, t) & = &
    \int\limits_{E_{\rm min}}^{+\infty} g(E - \bar{E})\, \alpha^{(1)}\, e^{ia_{2}a -iEt/\hbar}\; dE &
    \mbox{at } R_{1}<a<R_{2}, \\
  \psi^{(1)}_{\rm ref}(a, t) & = &
    \int\limits_{E_{\rm min}}^{+\infty} g(E - \bar{E})\, A_{R}^{(1)}\, e^{-ik_{1}a -iEt/\hbar}\; dE
        & \mbox{at } 0<a<R_{1},
\end{array}
\label{eq.app.4.2}
\end{equation}
where $k_{2} = \sqrt{E-V_{1}}$.
We find new unknown coefficients $\alpha^{(1)}$ and $A_{R}^{(1)}$, using requirements of continuity of the total wave function $\psi(a,t)$ (which is summation of all packets)
and its derivative at $R_{1}$:
\begin{equation}
\begin{array}{ll}
  \alpha^{(1)} = \displaystyle\frac{2\,k_{1}}{k_{1}+k_{2}}\, e^{i\,(k_{1}-k_{2})\,R_{1}}, &
  A_{R}^{(1)}  = \displaystyle\frac{k_{1}-k_{2}}{k_{1}+k_{2}}\, e^{2i\,k_{1}\,R_{1}}.
\end{array}
\label{eq.app.4.3}
\end{equation}

In the second step we consider further propagation of the packet $\psi^{(1)}_{\rm tr}(a, t)$, which is incident on the second boundary at $R_{2}$. It transforms into two new packets: the first packet transmitted through this boundary and propagating in the region III, and the second one reflected from this boundary and propagating back in the region II.
We define these packets in the form
\begin{equation}
\begin{array}{lcl}
  \psi^{(2)}(a, t) & = &
    \int\limits_{E_{\rm min}}^{+\infty} g(E - \bar{E})\: \varphi^{(2)}(a)\: e^{-iEt/\hbar}\; dE,
\end{array}
\label{eq.app.4.4}
\end{equation}
where as the stationary parts we use:
\begin{equation}
\begin{array}{lcll}
\varphi_{\rm inc}^{(2)}(a) & = & \alpha^{(1)} e^{ik_{2}a},
        & \mbox{for } R_{1}<a<R_{2}, \\
\varphi_{\rm tr}^{(2)}(a) & = & A_{T}^{(1)}e^{ika},
        & \mbox{for } a>R_{2}, \\
\varphi_{\rm ref}^{(2)}(a) & = & \beta^{(1)} e^{-ik_{2}a},
        & \mbox{for } R_{1}<a<R_{2},
\end{array}
\label{eq.app.4.5}
\end{equation}
where $k = \sqrt{E}$. Imposing condition of continuity on the total wave function and its derivative at $R_{2}$, we obtain two new equations,
from which we find new unknowns coefficients $A_{T}^{(1)}$ and $\beta^{(1)}$:
\begin{equation}
\begin{array}{ll}
  A_{T}^{(1)} = T_{2}^{+} \cdot \alpha^{(1)}, &
  T_{2}^{+} = \displaystyle\frac{2\,k_{2}}{k_{2}+k}\, e^{i\,(k_{2}-k)\,R_{2}}, \\
  \beta^{(1)} = R_{2}^{+} \cdot \alpha^{(1)}, &
  R_{2}^{+} = \displaystyle\frac{k_{2}-k}{k_{2}+k}\, e^{2i\,k_{2}\,R_{2}}.
\end{array}
\label{eq.app.4.6}
\end{equation}
We have introduced two new coefficients $T_{2}^{+}$, $R_{2}^{+}$, which logically connect the transmitted and reflected amplitudes $A_{T}^{(1)}$ and $\beta^{(1)}$ with the incident amplitude $\alpha^{(1)}$ in this step. Here, we shall use bottom index for denotation of number of the considered boundary, upper (top) sign ``$+$'' or ``$-$'' for positive (to the right) or negative (to the left) direction of the incident wave, correspondingly.
So, we can write $T_{1}^{+} = \alpha^{(1)}$ and $R_{1}^{+} = A_{R}^{(1)}$ also.

In the third step we consider further propagation of the reflected packet $\psi^{(2)}_{\rm ref}$ in the region II. Incidenting on the first boundary, it transforms into new packet $\psi^{(3)}_{\rm tr}$, transmitted through this boundary and propagating in the region I, and into new packet $\psi^{(3)}_{\rm ref}$, reflected from boundary and propagating back in the region II. We define the new packets by eq.~(\ref{eq.app.4.4}) (with upper index 3),
where as the stationary parts we use:
\begin{equation}
\begin{array}{lcll}
  \varphi_{\rm inc}^{(3)}(a) & = & \varphi_{\rm ref}^{(2)}(a),
          & \mbox{for } R_{1}<a<R_{2}, \\
  \varphi_{\rm tr}^{(3)}(a) & = & A_{R}^{(2)}e^{-ik_{1}a},
          & \mbox{for } 0<a<R_{1}, \\
  \varphi_{\rm ref}^{(3)}(a) & = & \alpha^{(2)} e^{ik_{2}a},
          & \mbox{for } R_{1}<a<R_{2}.
\end{array}
\label{eq.app.4.7}
\end{equation}
From continuity conditions for the total wave function and its derivative at $R_{1}$, we find the unknowns coefficients
$A_{R}^{(2)}$ and $\alpha^{(2)}$:
\begin{equation}
\begin{array}{ll}
  A_{T}^{(2)}  = T_{1}^{-} \cdot \beta^{(1)}, &
  T_{1}^{-} = \displaystyle\frac{2\,k_{2}}{k_{1}+k_{2}}\, e^{i\,(k_{1}-k_{2})\,R_{1}}, \\
  \alpha^{(2)} = R_{1}^{-} \cdot \beta^{(1)}, &
  R_{1}^{-} = \displaystyle\frac{k_{2}-k_{1}}{k_{1}+k_{2}}\, e^{-2i\,k_{2}\,R_{1}}.
\end{array}
\label{eq.app.4.8}
\end{equation}

In the forth step we need to consider further propagation of the reflected packet $\psi_{\rm ref}^{(1)}$ in the region I in the 1-st step. It is incident on the first boundary at $a=0$ transforming into new packet propagated to the right. At such a point we can include different considerations of origin of possible sources at $a=0$, possible full propagation (like in spherically symmetric problems of quantum decay in nuclear physics which is 3-dimensional and we have no additional boundaries at $a=0$) or, in contrary, full reflection used in different fully quantum approaches (like introduction of an infinite potential wall at $a=0$ in Ref.~\cite{AcacioDeBarros.2007.PRD}).
In order to produce ability to work with different such considerations, we write:
\begin{equation}
\begin{array}{lll}
  \varphi_{\rm inc}^{(4)}(a) = \varphi_{\rm ref}^{(1)}(a), &
  \varphi_{\rm tr}^{(4)}(a, k_{1}) = R_{0}^{-} \cdot \varphi_{\rm inc}^{(4)}(a,-k_{1}) =
    A_{\rm ref}^{(4)}\,e^{ik_{1}a}, &
  \mbox{for } 0<a<R_{1},
\end{array}
\label{eq.app.4.9}
\end{equation}
where
\begin{equation}
  A_{\rm ref}^{(4)} = R_{0}^{-} \cdot A_{\rm inc}^{(1)}.
\label{eq.app.4.10}
\end{equation}
Supposing the full propagation through this boundary (without any possible reflections), we obtain $R_{0}^{-} = -1$.
If we liked to use the condition of the infinite potential wall at $a=0$, than we should also have $R_{0}^{-} = -1$.

Analyzing further reflections and transmission of the packets concerning the boundaries, we conclude that any of following steps is similar to one of 4 considered above. From analysis of these steps recurrent relations are found for calculation of new unknown amplitudes $A_{\rm inc}^{(n)}$, $A_{R}^{(n)}$, $A_{T}^{(n)}$ $\alpha^{(n)}$ and $\beta^{(n)}$ for arbitrary step $n$, summations of these amplitudes are calculated. These series can be calculated easier, using coefficients $T_{i}^{\pm}$ and $R_{i}^{\pm}$. Analyzing all possible ``paths''
of the propagations of all possible packets inside the barrier and internal well, we obtain:
\begin{equation}
\begin{array}{lcl}
  \sum\limits_{n=1}^{+\infty} A_{\rm inc}^{(n)} & = &
    1 + \tilde{R}_{1}^{+}\,R_{0}^{-} + \tilde{R}_{1}^{+}\,R_{0}^{-} \cdot \tilde{R}_{1}^{+}\,R_{0}^{-} + ... =
    1 + \sum\limits_{m=1}^{+\infty} \bigl(\tilde{R}_{1}^{+}\,R_{0}^{-}\bigr)^{m} =
    \displaystyle\frac{1}{1 - \tilde{R}_{1}^{+}\,R_{0}^{-}}, \\

  \sum\limits_{n=1}^{+\infty} A_{T}^{(n)} & = &
  \Bigl( \sum\limits_{n=1}^{+\infty} A_{\rm inc}^{(n)} \Bigr) \cdot
  \Bigl\{ T_{1}^{+}\,T_{2}^{+} + T_{1}^{+}\cdot R_{2}^{+}\,R_{1}^{-}\cdot T_{2}^{+} + ... \Bigr\} =
  \Bigl( \sum\limits_{n=1}^{+\infty} A_{\rm inc}^{(n)} \Bigr) \cdot \tilde{T}_{1}^{+}, \\

  \sum\limits_{n=1}^{+\infty} A_{R}^{(n)} & = &
    \tilde{R}_{1}^{+} + \tilde{R}_{1}^{+} \cdot R_{0}^{-}\,\tilde{R}_{1}^{+} +
    \tilde{R}_{1}^{+} \cdot R_{0}^{-}\,\tilde{R}_{1}^{+} \cdot R_{0}^{-}\,\tilde{R}_{1}^{+} + ... = \\
\vspace{2mm}
  & = &
    \tilde{R}_{1}^{+} \cdot
      \Bigl( 1 + \sum\limits_{m=1}^{+\infty} \bigl(R_{0}^{-}\,\tilde{R}_{1}^{+}\bigr)^{m} \Bigr) =
    \displaystyle\frac{\tilde{R}_{1}^{+}} {1 - R_{0}^{-}\,\tilde{R}_{1}^{+}} =
  \Bigl( \sum\limits_{n=1}^{+\infty} A_{\rm inc}^{(n)} \Bigr) \cdot \tilde{R}_{1}^{+},
\end{array}
\label{eq.app.4.11}
\end{equation}
where
\begin{equation}
\begin{array}{lcl}
  \tilde{R}_{1}^{+} & = &
    R_{1}^{+} + T_{1}^{+}\,R_{2}^{+}\,T_{1}^{-} \cdot
      \Bigl( 1 + \sum\limits_{m=1}^{+\infty} \bigl(R_{2}^{+}\,R_{1}^{-}\bigr)^{m} \Bigr) =
    R_{1}^{+} + \displaystyle\frac{T_{1}^{+}\,R_{2}^{+}\,T_{1}^{-}} {1 - R_{2}^{+}\,R_{1}^{-}}, \\

  \tilde{T}_{1}^{+} & = &
    T_{1}^{+}\, T_{2}^{+} \cdot
    \Bigl( 1 + \sum\limits_{m=1}^{+\infty} \bigl(R_{2}^{+}\,R_{1}^{-}\bigr)^{m} \Bigr) =
    \displaystyle\frac{T_{1}^{+}\, T_{2}^{+}}{1 - R_{2}^{+}\,\tilde{R}_{1}^{-}}.
\end{array}
\label{eq.app.4.12}
\end{equation}

%
%

The resultant expressions for the incident, transmitted and reflected packets concerning the barrier are written in form of eq.~(\ref{eq.app.4.4}),
where the following stationary wave functions should be used:
\begin{equation}
\begin{array}{lcll}
\varphi_{\rm inc}(a) & = & e^{ik_{1}a},
                        & \mbox{for } 0<a<R_{1}, \\
\varphi_{\rm tr}(a)  & = & \sum\limits_{n=0}^{+\infty} A_{T}^{n} e^{ika},
                        & \mbox{for } a>R_{2}, \\
\varphi_{\rm ref}(a) & = & \sum\limits_{n=0}^{+\infty} A_{R}^{n} e^{-ik_{1}a},
                        & \mbox{for } 0<a<R_{1}.
\end{array}
\label{eq.app.4.13}
\end{equation}
At finishing, we determine the full amplitudes
\begin{equation}
\begin{array}{cccc}
  A_{T} = \sum\limits_{n=1}^{+\infty} A_{T}^{(n)}, &
  A_{R} = \sum\limits_{n=1}^{+\infty} A_{R}^{(n)}, &
  \alpha = \sum\limits_{n=1}^{+\infty} \alpha^{(n)} = \displaystyle\frac{A_{T}}{T_{2}^{+}}, &
  \beta = \sum\limits_{n=1}^{+\infty} \beta^{(n)} = \alpha \cdot R_{2}^{+}
\end{array}
\label{eq.app.4.14}
\end{equation}
and coefficients $T$ and $R$ describing penetration of the packet from the internal region outside
and its reflection from the barrier
\begin{equation}
\begin{array}{ll}
\vspace{1mm}
  T_{MIR} \equiv \displaystyle\frac{k}{k_{1}}\; \bigl|A_{T}\bigr|^{2} =
    \bigl|A_{\rm inc}\bigr|^{2} \cdot T_{\rm bar}, &
  T_{\rm bar} = \displaystyle\frac{k}{k_{1}}\; \bigl|\tilde{T}_{1}^{+} \bigr|^{2}, \\
  R_{MIR} \equiv \bigl|A_{R}\bigr|^{2} = \bigl|A_{\rm inc}\bigr|^{2} \cdot R_{\rm bar}, &
  R_{\rm bar} = \bigl|\tilde{R}_{1}^{+} \bigr|^{2},
\end{array}
\label{eq.app.4.15}
\end{equation}
where $T_{\rm bar}$ and $R_{\rm bar}$ are coefficients of penetrability and reflection of the barrier (in standard definition), and $\bigl|A_{\rm inc}\bigr|^{2}$ is coefficient determining oscillations of the packet inside the internal region (this is fully quantum analog of the normalization factor F introduced in Ref.~\cite{Gurvitz.1987.PRL} for semiclassical description of nuclear decay).

Series $\sum A_{\rm inc}^{(n)}$, $\sum A_{T}^{(n)}$, $\sum A_{R}^{(n)}$, $\sum \alpha^{(n)}$ and $\sum \beta^{(n)}$ obtained using the approach of the multiple internal reflections, \underline{exactly} coincide with the corresponding coefficients $A_{\rm inc}$, $A_{T}$, $A_{R}$, $\alpha$ and $\beta$ calculated by standard stationary method (where the continuity conditions of the stationary total wave function and its derivative are used at each boundaries, and the wave function is not equal to zero at $a=0$).
We test property:
\begin{equation}
\begin{array}{ccc}
  \displaystyle\frac{k}{k_{1}}\; |A_{T}|^{2} + |A_{R}|^{2} = 1 & \mbox{ or }&
  T_{MIR} + R_{MIR} = 1,
\end{array}
\label{eq.app.4.16}
\end{equation}
which is fulfilled and confirms that the method MIR gives us proper solution for the wave function.
If energy is less then the height of the barrier, then for description of penetration of the wave through such a barrier with its tunneling it needs to use 
the following change~\cite{Maydanyuk.2002.JPS,Maydanyuk.2003.PhD-thesis,Maydanyuk.2006.FPL}:
\begin{equation}
\begin{array}{cc}
  k_{2} \to i\,\xi, &
  \xi = \sqrt{E-V_{1}}.
\end{array}
\label{eq.app.4.17}
\end{equation}
Using it, all found above solutions are applied for the problem with tunneling through the barrier. For the barrier consisting from two rectangular steps of arbitrary heights and widths we have already obtained exact coincidence between amplitudes for the wave function obtained by method of MIR and the corresponding amplitudes obtained by standard approach of quantum mechanics. Increasing of number of the rectangular steps in the barriers keeps such a coincidence and fulfillment of the property~(\ref{eq.app.4.16}) (see some generalizations in Refs.~\cite{Maydanyuk.2003.PhD-thesis,Maydanyuk.arXiv:0805.4165,Maydanyuk.arXiv:0906.4739}). So, we have obtained full coincidence between all amplitudes calculated by method MIR and by standard approach of quantum mechanics, and that is way we generalize the method MIR for description of penetration of the wave through barrier consisting from arbitrary number of rectangular steps of arbitrary sizes.


\section{Calculations of the penetrability}

The penetrabilities of the barrier with $A=36$, $B=12\,\Lambda$ parameters at $\Lambda=0.01$ for the closed Friedmann--Robertson--Walker model with quantization in the presence of a positive cosmological constant and radiation calculated by the fully quantum approach and by the semiclassical approach are presented in Tabl.~1. One can see that inside the whole region of the energy of radiation $E_{\rm rad}$ up to 2690 the penetrability calculated by the fully quantum method is very close to its value, obtained by the semiclassical approach by eqs.~(\ref{eq.6.1}) and (\ref{eq.6.2}). Also one can find an essential difference between these calculations and results obtained in Ref.~\cite{AcacioDeBarros.2007.PRD}
by the non-stationary quantum approach.
\begin{table}
\hspace{-20mm}
\begin{center}
\begin{tabular}{|c|c|c|c|c|c|c|} \hline
  Energy
  & \multicolumn{2}{|c|}{Penetrability $P_{\rm penetrability}$}
  & \multicolumn{2}{|c|}{Time $\tau$} & \multicolumn{2}{|c|}{Turning points} \\ \cline{2-7}
  $E_{\rm rad}$
  & Fully quantum method & Method WKB & Fully quantum method & Method WKB & $a_{\rm tp,\, in}$ & $a_{\rm tp,\, out}$ \\ \hline
   1.0 & $8.7126 \times 10^{-521}$ & $2.0888 \times 10^{-521}$ & $3.8260 \times 10^{+519}$ & $1.5958 \times 10^{+520}$ & 0.16 &  17.31 \\
   2.0 & $2.4225 \times 10^{-520}$ & $5.5173 \times 10^{-521}$ & $1.9460 \times 10^{+519}$ & $8.5448 \times 10^{+519}$ & 0.23 &  17.31 \\
   3.0 & $6.2857 \times 10^{-520}$ & $1.3972 \times 10^{-520}$ & $9.1863 \times 10^{+518}$ & $4.1326 \times 10^{+519}$ & 0.28 &  17.31 \\
   4.0 & $1.5800 \times 10^{-519}$ & $3.4428 \times 10^{-520}$ & $4.2201 \times 10^{+518}$ & $1.9367 \times 10^{+519}$ & 0.33 &  17.31 \\
   5.0 & $3.8444 \times 10^{-519}$ & $8.2935 \times 10^{-520}$ & $1.9392 \times 10^{+518}$ & $8.9892 \times 10^{+518}$ &  0.37 &  17.31 \\
   6.0 & $9.2441 \times 10^{-519}$ & $1.9701 \times 10^{-519}$ & $8.8350 \times 10^{+517}$ & $4.1455 \times 10^{+518}$ &  0.40 &  17.31 \\
   7.0 & $2.1678 \times 10^{-518}$ & $4.5987 \times 10^{-519}$ & $4.0694 \times 10^{+517}$ & $1.9183 \times 10^{+518}$ &  0.44 &  17.31 \\
   8.0 & $5.0192 \times 10^{-518}$ & $1.0621 \times 10^{-518}$ & $1.8790 \times 10^{+517}$ & $8.8797 \times 10^{+517}$ &  0.47 &  17.31 \\
   9.0 & $1.1604 \times 10^{-517}$ & $2.4316 \times 10^{-518}$ & $8.6212 \times 10^{+516}$ & $4.1140 \times 10^{+517}$ &  0.50 &  17.31 \\
  10.0 & $2.6279 \times 10^{-517}$ & $5.5016 \times 10^{-518}$ & $4.0128 \times 10^{+516}$ & $1.9168 \times 10^{+517}$ &  0.52 &  17.31 \\

100.0 & $1.6165 \times 10^{-490}$ & $3.1959 \times 10^{-491}$ & $2.0717 \times 10^{+490}$ & $ 1.0478 \times 10^{+491}$ & 1.67 & 17.23 \\
200.0 & $ 8.5909 \times 10^{-465}$ & $1.6936 \times 10^{-465}$ & $5.5397 \times 10^{+464}$ & $ 2.8100 \times 10^{+465}$ & 2.37 & 17.15 \\
300.0 & $ 6.8543 \times 10^{-441}$ & $1.3419 \times 10^{-441}$ & $8.5461 \times 10^{+440}$ & $ 4.3653 \times 10^{+441}$ & 2.92 & 17.07 \\
400.0 & $ 3.6688 \times 10^{-418}$ & $7.1642 \times 10^{-419}$ & $1.8531 \times 10^{+418}$ & $ 9.4900 \times 10^{+418}$ & 3.39 & 16.98 \\
500.0 & $ 2.6805 \times 10^{-396}$ & $5.2521 \times 10^{-397}$ & $2.8508 \times 10^{+396}$ & $ 1.4550 \times 10^{+397}$ & 3.82 & 16.89 \\
600.0 & $ 4.1386 \times 10^{-375}$ & $8.0511 \times 10^{-376}$ & $2.0338 \times 10^{+375}$ & $ 1.0454 \times 10^{+376}$ & 4.20 & 16.80 \\
700.0 & $ 1.7314 \times 10^{-354}$ & $3.3810 \times 10^{-355}$ & $5.2806 \times 10^{+354}$ & $ 2.7043 \times 10^{+355}$ & 4.57 & 16.70 \\
800.0 & $ 2.4308 \times 10^{-334}$ & $4.7497 \times 10^{-335}$ & $4.0448 \times 10^{+334}$ & $ 2.0701 \times 10^{+335}$ & 4.91 & 16.60 \\
900.0 & $ 1.3213 \times 10^{-314}$ & $2.5761 \times 10^{-315}$ & $7.9408 \times 10^{+314}$ & $ 4.0730 \times 10^{+315}$ & 5.24 & 16.50 \\
1000.0 & $ 3.0920 \times 10^{-295}$ & $6.0272 \times 10^{-296}$ & $3.5999 \times 10^{+295}$ & $ 1.8468 \times 10^{+296}$ & 5.56 & 16.40 \\
1100.0 & $ 3.4274 \times 10^{-276}$ & $6.6576 \times 10^{-277}$ & $3.4289 \times 10^{+276}$ & $ 1.7652 \times 10^{+277}$ & 5.87 & 16.29 \\
1200.0 & $ 1.9147 \times 10^{-257}$ & $3.7259 \times 10^{-258}$ & $6.4553 \times 10^{+257}$ & $ 3.3174 \times 10^{+258}$ & 6.18 & 16.18 \\
1300.0 & $ 5.8026 \times 10^{-239}$ & $1.1253 \times 10^{-239}$ & $2.2333 \times 10^{+239}$ & $ 1.1516 \times 10^{+240}$ & 6.47 & 16.06 \\
1400.0 & $ 9.9042 \times 10^{-221}$ & $1.9252 \times 10^{-221}$ & $1.3683 \times 10^{+221}$ & $ 7.0393 \times 10^{+221}$ & 6.77 & 15.93 \\
1500.0 & $ 1.0126 \times 10^{-202}$ & $1.9551 \times 10^{-203}$ & $1.3965 \times 10^{+203}$ & $ 7.2333 \times 10^{+203}$ & 7.07 & 15.81 \\
1600.0 & $ 6.2741 \times 10^{-185}$ & $1.2155 \times 10^{-185}$ & $2.3480 \times 10^{+185}$ & $ 1.2119 \times 10^{+186}$ & 7.36 & 15.67 \\
1700.0 & $ 2.4923 \times 10^{-167}$ & $4.8143 \times 10^{-168}$ & $6.1488 \times 10^{+167}$ & $ 3.1831 \times 10^{+168}$ & 7.66 & 15.53 \\
1800.0 & $ 6.4255 \times 10^{-150}$ & $1.2437 \times 10^{-150}$ & $2.4783 \times 10^{+150}$ & $ 1.2803 \times 10^{+151}$ & 7.96 & 15.38 \\
1900.0 & $ 1.1189 \times 10^{-132}$ & $2.1580 \times 10^{-133}$ & $1.4776 \times 10^{+133}$ & $ 7.6619 \times 10^{+133}$ & 8.26 & 15.22 \\
2000.0 & $ 1.3288 \times 10^{-115}$ & $2.5653 \times 10^{-116}$ & $1.2914 \times 10^{+116}$ & $ 6.6895 \times 10^{+116}$ & 8.58 & 15.04 \\
2100.0 & $ 1.1105 \times 10^{-98}$  & $2.1357 \times 10^{-99}$  & $1.6036 \times 10^{+99}$ & $ 8.3382 \times 10^{+99}$ & 8.90 & 14.85 \\
2200.0 & $6.6054 \times 10^{-82}$ & $1.2690 \times 10^{-82}$ & $2.7988 \times 10^{+82}$ & $ 1.4567 \times 10^{+83}$ & 9.24 & 14.64 \\
2300.0 & $2.8693 \times 10^{-65}$ & $5.4647 \times 10^{-66}$ & $6.6952 \times 10^{+65}$ & $ 3.5154 \times 10^{+66}$ & 9.60 & 14.41 \\
2400.0 & $9.1077 \times 10^{-49}$ & $1.7297 \times 10^{-49}$ & $2.1959 \times 10^{+49}$ & $ 1.1562 \times 10^{+50}$ & 10.00 &  14.14 \\
2500.0 & $2.1702 \times 10^{-32}$ & $4.0896 \times 10^{-33}$ & $9.6290 \times 10^{+32}$ & $5.1098 \times 10^{+33}$ & 10.44 & 13.81 \\
2600.0 & $3.9788 \times 10^{-16}$ & $7.3137 \times 10^{-17}$ & $5.5322 \times 10^{+16}$ & $3.0096 \times 10^{+17}$ & 11.00 & 13.37 \\

2610.0 & $1.6663 \times 10^{-14}$ & $3.0428 \times 10^{-15}$ & $1.3290 \times 10^{+15}$ & $7.2780 \times 10^{+15}$ & 11.07 & 13.31 \\
2620.0 & $6.9240 \times 10^{-13}$ & $1.2606 \times 10^{-13}$ & $3.2187 \times 10^{+13}$ & $1.7678 \times 10^{+14}$ & 11.14 & 13.25 \\
2630.0 & $2.8842 \times 10^{-11}$ & $5.2116 \times 10^{-12}$ & $7.7789 \times 10^{+11}$ & $4.3050 \times 10^{+12}$ & 11.21 & 13.19 \\
2640.0 & $1.2002 \times 10^{-9}$ & $2.1495 \times 10^{-10}$ & $1.8825 \times 10^{+10}$ & $1.0511 \times 10^{+11}$ & 11.29 & 13.12 \\
2650.0 & $4.9881 \times 10^{-8}$ & $8.8401 \times 10^{-9}$ & $4.5642 \times 10^{+8}$ & $2.5754 \times 10^{+9}$ & 11.38 & 13.05 \\
2660.0 & $2.0738 \times 10^{-6}$ & $3.6263 \times 10^{-7}$ & $1.1068 \times 10^{+7}$ & $6.3303 \times 10^{+7}$ & 11.47 & 12.97 \\
2670.0 & $8.7110 \times 10^{-5}$ & $1.4836 \times 10^{-5}$ & $2.6596 \times 10^{+5}$ & $1.5615 \times 10^{+6}$ & 11.58 & 12.87 \\
2680.0 & $3.6953 \times 10^{-3}$ & $6.0519 \times 10^{-4}$ & $6.3369 \times 10^{+3}$ & $3.8693 \times 10^{+4}$ & 11.70 & 12.76 \\
2690.0 & $1.5521 \times 10^{-1}$ & $2.4634 \times 10^{-2}$ & $1.5293 \times 10^{+2}$ & $9.3602 \times 10^{+2}$ & 11.86 & 12.61 \\
\hline
\end{tabular}
\end{center}
\caption{\small The penetrability $P_{\rm penetrability}$ of the barrier and the duration $\tau$ of the formation of the Universe defined by eq.~(\ref{eq.6.3}) in the fully quantum and semiclassical approaches
\label{table.1}}
\end{table}
In the next Tabl.~2 the coefficients of the penetrability, reflection and mixing calculated in the fully quantum method are presented for the energy of radiation $E_{\rm rad}$ close to the height of the barrier. One can see that summation of all such values for coefficients allows to reconstruct the property (\ref{eq.4.2.3})
with accuracy of the first 11--18 digits.
\begin{table}
\hspace{-20mm}
\begin{center}
\begin{tabular}{|c|c|c|c|c|c|c|c|} \hline
  Energy & \multicolumn{4}{|c|}{Fully quantum method} &
  \multicolumn{2}{|c|}{Turning points} \\ \cline{2-7}
  $E_{\rm rad}$ &
  Penetrability &
  Reflection &
  Mixing &
  Summation &
  $a_{\rm tp,\, in}$ & $a_{\rm tp,\, out}$ \\ \hline
2690.0 & 0.15521440329121 & 0.84478559670782 & $1.47 \times 10^{-19}$ & 0.99999999999904 & 11.86 & 12.61 \\
2691.0 & 0.22040333134216 & 0.77959666865655 & $1.59 \times 10^{-19}$ & 0.99999999999871 & 11.88 & 12.59 \\
2692.0 & 0.30886886816339 & 0.69113113183491 & $7.61 \times 10^{-20}$ & 0.99999999999831 & 11.90 & 12.57 \\
2693.0 & 0.42120001498898 & 0.57879998500886 & $1.41 \times 10^{-19}$ & 0.99999999999785 & 11.93 & 12.55 \\
2694.0 & 0.55773509442073 & 0.44226490557664 & $1.44 \times 10^{-19}$ & 0.99999999999738 & 11.95 & 12.53 \\
2695.0 & 0.70351298662967 & 0.29648701336733 & $7.82 \times 10^{-20}$ & 0.99999999999701 & 11.98 & 12.50 \\
2696.0 & 0.84382355425692 & 0.15617644573996 & $7.80 \times 10^{-20}$ & 0.99999999999688 & 12.00 & 12.48 \\
2697.0 & 0.94803705920675 & 0.05196294079035 & $3.45 \times 10^{-20}$ & 0.99999999999711 & 12.04 & 12.44 \\
2698.0 & 0.99768097743782 & 0.00231902255993 & $2.10 \times 10^{-20}$ & 0.99999999999776 & 12.07 & 12.41 \\
2699.0 & 0.98255293343537 & 0.01744706656362 & $1.43 \times 10^{-20}$ & 0.99999999999901 & 12.12 & 12.36 \\
\hline
\end{tabular}
\end{center}
\caption{\small The coefficients of the penetrability, reflection and mixing calculated by the fully quantum method and their summation
\label{table.2}}
\end{table}

\begin{table}
\hspace{-20mm}
\begin{center}
\begin{tabular}{|c|c|c|c|c|c|} \hline
  Energy
  & \multicolumn{3}{|c|}{Penetrability $P_{\rm penetrability}$}
  & \multicolumn{2}{|c|}{Time $\tau$} \\ \cline{2-6}
  & Full QM method 1 & Full QM method 2 & Method WKB & Full QM method 1 & Method WKB \\ \hline
 10.0 &  $8.1070 \times 10^{-33}$ & $7.6149 \times 10^{-31}$ & $1.4522 \times 10^{-31}$ &  $8.6156 \times 10^{+32}$ &  $4.8094 \times 10^{+31}$ \\
 20.0 &  $7.6221 \times 10^{-30}$ & $7.7349 \times 10^{-29}$ & $1.4692 \times 10^{-29}$ &  $9.4313 \times 10^{+29}$ &  $4.8928 \times 10^{+29}$ \\
 30.0 &  $7.6975 \times 10^{-29}$ & $7.3089 \times 10^{-27}$ & $1.3848 \times 10^{-27}$ &  $9.5988 \times 10^{+28}$ &  $5.3354 \times 10^{+27}$ \\
 40.0 &  $3.5680 \times 10^{-26}$ & $6.5169 \times 10^{-25}$ & $1.2298 \times 10^{-25}$ &  $2.1257 \times 10^{+26}$ &  $6.1670 \times 10^{+25}$ \\
 50.0 &  $5.5831 \times 10^{-25}$ & $5.4707 \times 10^{-23}$ & $1.0285 \times 10^{-23}$ &  $1.3936 \times 10^{+25}$ &  $7.5647 \times 10^{+23}$ \\
 60.0 &  $2.0591 \times 10^{-22}$ & $4.3423 \times 10^{-21}$ & $8.1523 \times 10^{-22}$ &  $3.8719 \times 10^{+22}$ &  $9.7797 \times 10^{+21}$ \\
 70.0 &  $3.0663 \times 10^{-21}$ & $3.3043 \times 10^{-19}$ & $6.1642 \times 10^{-20}$ &  $2.6640 \times 10^{+21}$ &  $1.3251 \times 10^{+20}$ \\
 80.0 &  $1.5530 \times 10^{-18}$ & $2.3850 \times 10^{-17}$ & $4.4346 \times 10^{-18}$ &  $5.3862 \times 10^{+18}$ &  $1.8862 \times 10^{+18}$ \\
 90.0 &  $1.3181 \times 10^{-17}$ & $1.6564 \times 10^{-15}$ & $3.0658 \times 10^{-16}$ &  $6.4948 \times 10^{+17}$ &  $2.7923 \times 10^{+16}$ \\
100.0 &  $3.2922 \times 10^{-14}$ & $1.1053 \times 10^{-13}$ & $2.0304 \times 10^{-14}$ &  $2.6622 \times 10^{+14}$ &  $4.3167 \times 10^{+14}$ \\
110.0 &  $4.9414 \times 10^{-14}$ & $7.0911 \times 10^{-12}$ & $1.2935 \times 10^{-12}$ &  $1.8158 \times 10^{+14}$ &  $6.9367 \times 10^{+12}$ \\
120.0 &  $8.6052 \times 10^{-11}$ & $4.4005 \times 10^{-10}$ & $7.9523 \times 10^{-11}$ &  $1.0678 \times 10^{+11}$ &  $1.1555 \times 10^{+11}$ \\
130.0 &  $2.1009 \times 10^{-10}$ & $2.6431 \times 10^{-8}$ & $4.7194 \times 10^{-9}$ &  $4.4822 \times 10^{+10}$ &  $1.9953 \times 10^{+9}$ \\
140.0 &  $2.2012 \times 10^{-8}$ & $1.5460 \times 10^{-6}$ & $2.7128 \times 10^{-7}$ &  $4.3888 \times 10^{+8}$ &  $3.5612 \times 10^{+7}$ \\
150.0 &  $2.8361 \times 10^{-6}$ & $8.8293 \times 10^{-5}$ & $1.5114 \times 10^{-5}$ &  $3.4994 \times 10^{+6}$ &  $6.5663 \times 10^{+5}$ \\
160.0 &  $2.9685 \times 10^{-5}$ & $4.9980 \times 10^{-3}$ & $8.1663 \times 10^{-4}$ &  $3.4471 \times 10^{+5}$ &  $1.2530 \times 10^{+4}$ \\
170.0 &  $3.4894 \times 10^{-3}$ & $2.6078 \times 10^{-1}$ & $4.2919 \times 10^{-2}$ &  $3.0460 \times 10^{+3}$ &  $2.4820 \times 10^{+2}$ \\
\hline
\end{tabular}
\end{center}
\caption{\small
The penetrability $P_{\rm penetrability}$ of the barrier and duration $\tau$ of the formation of the Universe defined by eq.~(\ref{eq.6.3}) in the FRW-model with the Chaplygin gas obtained in the fully quantum and semiclassical approaches
(minimum of the hole is -93.579 and its coordinate is 1.6262,
maximum of the barrier is 177.99 and its coordinate is 5.6866):
the fully QM method 1 is calculations by the fully quantum approach for the boundary located in the coordinate of the minimum of the internal hole (i.~e. coordinate is 1.6262),
the fully QM method 2 is calculations by the fully quantum approach for the boundary located in the internal turning point $a_{\rm tp,\, in}$ (coordinates of the turning points are in Tabl.~4)
\label{table.3}}
\end{table}

\begin{table}
\hspace{-20mm}
\begin{center}
\begin{tabular}{|c|c|c|c|c|c|c|c|} \hline
  Energy & \multicolumn{4}{|c|}{Fully quantum method} &
  \multicolumn{2}{|c|}{Turning points} \\ \cline{2-7} &
  Penetrability &
  Reflection &
  Interference &
  Summation &
  $a_{\rm tp,\, in}$ & $a_{\rm tp,\, out}$ \\ \hline
 10.0 &  $8.1070216824 \times 10^{-33}$ &  1.00000000000000 &  $3.06 \times 10^{-20}$ &  1.00000000000000 &  3.49 &  7.08 \\
 20.0 &  $7.6221543404 \times 10^{-30}$ &  1.00000000000000 &  $8.55 \times 10^{-20}$ &  1.00000000000000 &  3.59 &  7.05 \\
 30.0 &  $7.6975296835 \times 10^{-29}$ &  1.00000000000000 &  $1.82 \times 10^{-20}$ &  1.00000000000000 &  3.69 &  7.01 \\
 40.0 &  $3.5680158760 \times 10^{-26}$ &  1.00000000000000 &  $2.34 \times 10^{-19}$ &  1.00000000000000 &  3.79 &  6.97 \\
 50.0 &  $5.5831154210 \times 10^{-25}$ &  1.00000000000000 &  $4.98 \times 10^{-20}$ &  1.00000000000000 &  3.89 &  6.92 \\
 60.0 &  $2.0591415452 \times 10^{-22}$ &  1.00000000000000 &  $4.86 \times 10^{-20}$ &  1.00000000000000 &  3.98 &  6.88 \\
 70.0 &  $3.0663252971 \times 10^{-21}$ &  1.00000000000000 &  $1.84 \times 10^{-19}$ &  1.00000000000000 &  4.08 &  6.83 \\
 80.0 &  $1.5530040238 \times 10^{-18}$ &  1.00000000000000 &  $2.08 \times 10^{-19}$ &  1.00000000000000 &  4.18 &  6.78 \\
 90.0 &  $1.3181086626 \times 10^{-17}$ &  1.00000000000000 &  $5.03 \times 10^{-20}$ &  1.00000000000000 &  4.28 &  6.73 \\
100.0 &  $3.2922846164 \times 10^{-14}$ &  0.99999999999996 &  $3.13 \times 10^{-20}$ &  1.00000000000000 &  4.38 &  6.67 \\
110.0 &  $4.9414392175 \times 10^{-14}$ &  0.99999999999995 &  $1.45 \times 10^{-19}$ &  1.00000000000000 &  4.48 &  6.61 \\
120.0 &  $8.6052092530 \times 10^{-11}$ &  0.99999999991394 &  $1.06 \times 10^{-19}$ &  1.00000000000000 &  4.59 &  6.55 \\
130.0 &  $2.1009662247 \times 10^{-10}$ &  0.99999999978990 &  $2.68 \times 10^{-20}$ &  1.00000000000000 &  4.70 &  6.48 \\
140.0 &  $2.2012645564 \times 10^{-8}$ &  0.99999997798735 &  $1.60 \times 10^{-19}$ &  1.00000000000000 &  4.83 &  6.39 \\
150.0 &  $2.8361866579 \times 10^{-6}$ &  0.99999716381330 &  $4.89 \times 10^{-20}$ &  0.99999999999996 &  4.96 &  6.30 \\
160.0 &  $2.9685643504 \times 10^{-5}$ &  0.99997031435611 &  $6.94 \times 10^{-20}$ &  0.99999999999961 &  5.11 &  6.18 \\
170.0 &  $3.4894544195 \times 10^{-3}$ &  0.99651054553176 &  $2.02 \times 10^{-19}$ &  0.99999999995131 &  5.31 &  6.02 \\
\hline
\end{tabular}
\end{center}
\caption{\small The coefficients of the penetrability, reflection and mixing calculated by the fully quantum method and test on their summation for the FRW-model with the Chaplygin gas density component (the fully quantum approach 1 is used at the internal boundary located in the coordinate of the minimum of the internal hole)
\label{table.4}}
\end{table}




\end{document}